\documentclass[reprint,amsmath,amssymb,aps,superscriptaddress,floatfix]{revtex4-1}

\usepackage{graphicx}
\usepackage{dcolumn}
\usepackage{bm}
\usepackage{hyperref}
\usepackage{siunitx}
\usepackage{color}

\DeclareSIUnit\kT{$k_B T$}
\DeclareSIUnit\dyne{dyne}
\DeclareSIUnit\molar{M}
\newcommand{\bnabla}{\mbox{\boldmath$\nabla$}}
\newcommand{\bdiv}{\bnabla\cdot}
\newcommand{\bcurl}{\bnabla\times}
\newcommand{\dd}{\mathrm{d}}
\newcommand{\ee}{\mathrm{e}}
\newcommand{\ii}{\mathrm{i}}
\newcommand{\order}{\mathcal{O}}
\newcommand{\ve}[1]{\bm{\mathbf{#1}}}
\newcommand{\veh}[1]{\bm{\mathbf{\hat{#1}}}}

\newcommand{\kT}{{k_B T}}
\newcommand{\avg}[1]{\left\langle #1 \right\rangle}
\newcommand{\deldel}[2]{\frac{\delta #1}{\delta #2}}
\newcommand{\parpar}[2]{\frac{\partial #1}{\partial #2}}

\begin{document}

\title{Entropic forces stabilize diverse emergent structures in colloidal membranes}

\date{\today}

\author{Louis Kang}
\email{lkang@mail.med.upenn.edu}
\affiliation{Department of Physics \& Astronomy, University of Pennsylvania, 203 South 33rd Street, Philadelphia, Pennsylvania 19104, USA}

\author{Thomas Gibaud}
\affiliation{Laboratoire de Physique, \'{E}cole Normale Sup\'{e}rieure de Lyon, Universit\'{e} de Lyon, CNRS/UMR 5672, 46 all\'{e}e d'Italie, 69007 Lyon, France}

\author{Zvonimir Dogic}
\affiliation{The Martin Fisher School of Physics, Brandeis University, 415 South Street, Waltham, Massachusetts 02454, USA}

\author{T. C. Lubensky}
\affiliation{Department of Physics \& Astronomy, University of Pennsylvania, 203 South 33rd Street, Philadelphia, Pennsylvania 19104, USA}

\begin{abstract}
The depletion interaction mediated by non-adsorbing polymers promotes condensation and assembly of repulsive colloidal particles into diverse higher-order structures and materials. One example, with particularly rich emergent behaviors, is the formation of two-dimensional colloidal membranes from a suspension of filamentous \emph{fd} viruses, which act as rods with effective repulsive interactions, and dextran, which acts as a condensing, depletion-inducing agent. Colloidal membranes exhibit chiral twist even when the constituent virus mixture lacks macroscopic chirality, change from a circular shape to a striking starfish shape upon changing the chirality of constituent rods, and partially coalesce via domain walls through which the viruses twist by \SI{180}{\degree}. We formulate an entropically-motivated theory that can quantitatively explain these experimental structures and measurements, both previously published and newly performed, over a wide range of experimental conditions. Our results elucidate how entropy alone, manifested through the viruses as Frank elastic energy and through the depletants as an effective surface tension, drives the formation and behavior of these diverse structures. Our generalizable principles propose the existence of analogous effects in molecular membranes and can be exploited in the design of reconfigurable colloidal structures.
\end{abstract}

\maketitle

\section{Introduction}

Suspensions of particles with hard-core repulsive interactions form equilibrium phases that minimize the systems' free energy by maximizing their entropy. Since entropy is conventionally associated with disorder, it might be expected that hard-particle fluids form structures that lack long-range order. However, extensive experimental work and theoretical models have repeatedly demonstrated the counterintuitive notion that entropy alone is sufficient to stabilize ordered phases of ever-increasing complexity. Among other examples, it has been shown that entropy can drive formation of 3D bulk crystals in suspensions of hard spheres~\cite{Pusey:1986dv}, nematic and smectic liquid crystalline phases with hard rods~\cite{Onsager:1949wv,Frenkel:1988fv}, and more exotic binary crystals and diverse microphase-separated states in mixtures of hard particles~\cite{Eldridge:1993hx,Adams:1998fa}.  

Recent work has demonstrated that a mixture of monodisperse micron-long filamentous bacteriophages and non-adsorbing polymers assemble into 2D one-rod-length-thick colloidal monolayer membranes~\cite{Barry:2010cz,Yang:2012ds}. Colloidal membranes exhibit an exceedingly rich phenomenology. They support a myriad of defects including twist domain walls and linear arrays of pores~\cite{Zakhary:2014de}. Increasing chirality induces a transition of flat 2D membranes into 1D twisted ribbons, and mixing rods of multiple lengths leads to formation of finite-sized colloidal rafts that are evocative of similar structures observed in conventional  lipid bilayers~\cite{Gibaud:2012cf,Sharma:2014cl}. All of these complex mesoscopic behaviors arise from very simple microscopic interactions between constituent particles. Filamentous viruses interact only through an effective hard-rod repulsion. Similarly, the uncharged dextran molecules act as effective Asakura-Oosawa penetrable spheres~\cite{Asakura:1954jy,Asakura:1958kc}. From this perspective, the virus particles and dextran molecules comprise a gas of hard rods and hard spheres, and the structures found in colloidal membranes must be stabilized by entropic, hard-core interactions~\cite{Frenkel:1993ws}. We formulate a theoretical model based purely on such entropic considerations. Our model explains many known structural features of colloidal membranes and directly relates them to the known entropic interactions in rod/polymer mixtures. Furthermore, it makes a number of new predictions that are directly verified by new experimental results.  

Colloidal suspensions are a quintessential model system in soft condensed matter physics.  They are not only interesting in their own right but also provide new insights into the structure and dynamics of diverse phases; these insights only depend on the symmetries of the constituent particles and are thus relevant on all lengthscales. For example, engineering colloidal shapes and interactions makes it possible to mimic many processes found in atomic and molecular systems, including liquid-gas phase separation, wetting, thermal capillary waves, crystal nucleation, and the glass transition~\cite{Aarts:2004hi,Gasser:2001fo,Gast:1983if,Lekkerkerker:2007ij, Pusey:1986dv,Weeks:2000dw}. In stark contrast to molecular systems, the size of model colloids makes it is possible to directly track the positions of all the constituent particles, thus yielding important information about universal physical processes in various condensed matter systems. Conventional fluid membranes, assembled from permanently-linked hydrophobic and hydrophilic components, are another interesting and important soft matter system and play an essential role in biology~\cite{alberts}. However, due to our inability to directly visualize real-time dynamics of lipid bilayers at the nanometer scale, many membrane-based processes remain poorly understood. Intriguingly, the large-scale elastic deformations of colloidal membranes are described by the same continuum theories that are used to describe conventional lipid bilayers. Based on this observation and following the analogy between colloids and molecular substances, we hope that colloidal membranes will provide new understanding about universal membrane-mediated behaviors. There have been some recent overtures in this vein. For example, colloidal membranes permit direct visualization and quantitative characterization of liquid raft-like clusters~\cite{Sharma:2014cl}, a subject that remains controversial in conventional lipid membranes~\cite{Lingwood:2009kf,Sharma:2014cl}. Eventual understanding of such complex structures requires a theoretical model that relates mesoscopic properties of colloidal membranes to the microscopic interactions of their constituent building blocks. 

The rest of the paper is organized as follows. In Sec.~\ref{sec:over} we briefly review the rich phenomenology of colloidal membranes. In Sec.~\ref{sec:res}, we introduce a new entropy-based theoretical model of colloidal membranes and compare our results to known properties of colloidal membranes, including static edge fluctuation data [Figs.~\ref{fig:tension}(b), \ref{fig:tension}(d), and \ref{fig:tension}(e)] and twist domain wall retardance (Fig.~\ref{fig:wallret})~\cite{Gibaud:2012cf,Zakhary:2014de}. Furthermore, we also discuss new predictions of our theoretical model, including how the structure of the membrane's edge depends on membrane radius (Fig.~\ref{fig:edgeret}) and dynamical edge fluctuation data [Fig.~\ref{fig:tension}(c)]. These predictions are tested against new experimental data. Section~\ref{sec:th} explains the model in complete detail and Sec.~\ref{sec:exp} describes experimental methods. Finally, we summarize our findings and discuss their wider implications in Sec.~\ref{sec:dis}.

\section{\label{sec:over}Overview of colloidal membranes}

\begin{figure*}
	\includegraphics[width=\textwidth]{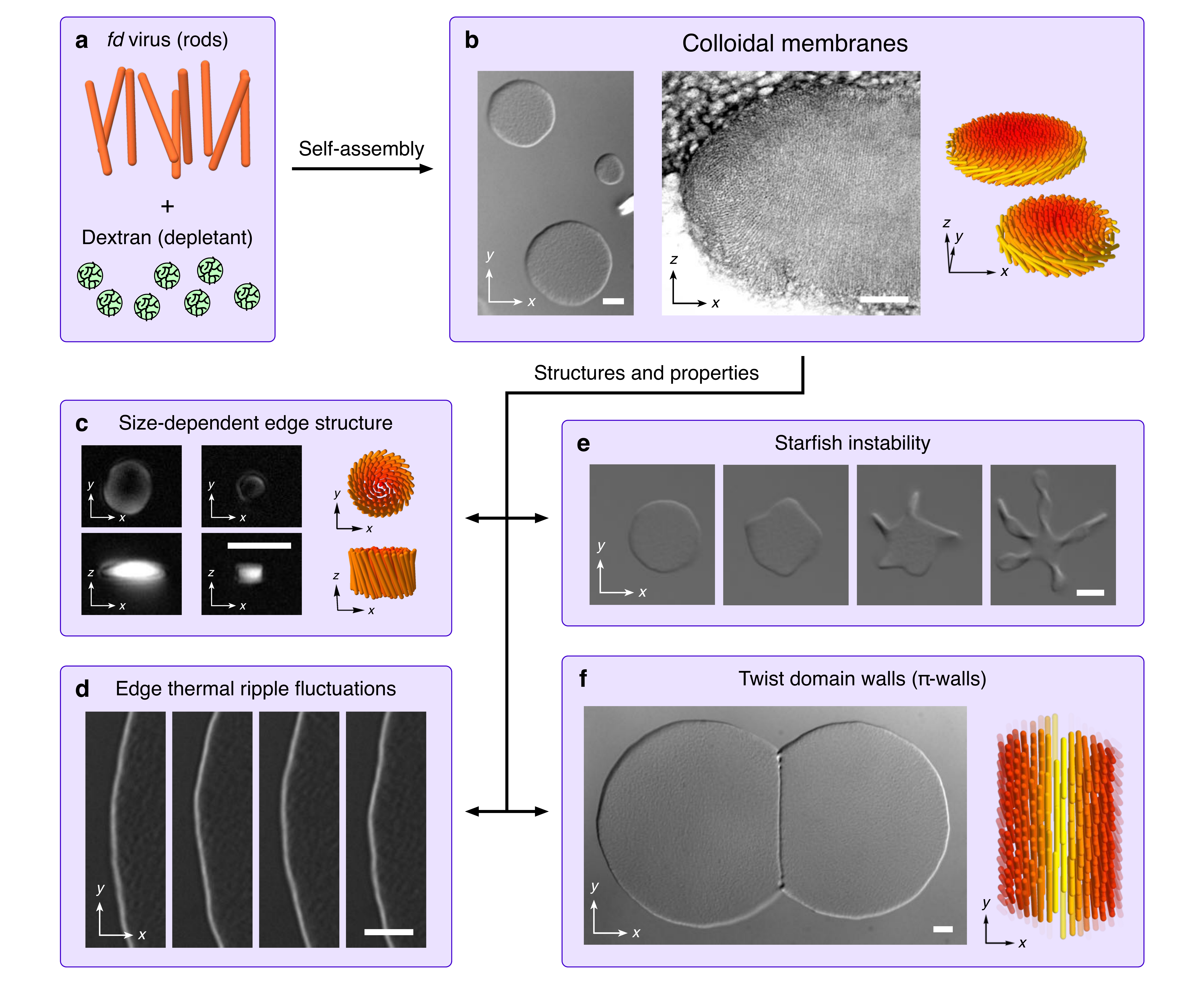}
	\caption{\label{fig:images}(Color) Overview of colloidal membranes. (a) \emph{fd} virus particles and dextran molecules act as rod-shaped colloids and spherical depletants, respectively. (b) Depleting molecules condense a dilute isotropic virus suspension into a liquid-like colloidal monolayer of aligned rods. From left to right, differential interference contrast (DIC) image of circular membranes of various sizes, transmission electron microscopy image showing a curved cross-section of the edge of a large membrane, and schematic of two large circular membranes of opposite chirality. (c) From left to right, top- and side-view LC-Polscope images of a medium-sized membrane, top- and side-view LC-Polscope images of a small membrane, and top- and side-view LC schematics of a small membrane. Along with (b), these images illustrate that edges of smaller membranes are more squared. (d) DIC images of thermally-excited ripple fluctuations at four different times. (e) DIC images of a temperature induced transition of a flat 2D colloidal membrane (left) into a structure with a starfish morphology (right). (f) DIC image (left) and schematic (right) of a twist domain wall, or $\pi$-wall, formed from two partially-coalesced circular membranes. (b) (left), (c), (d), (e), (f) Scale bars, \SI{4}{\um}. (b) (middle) Scale bar, \SI{0.2}{\um}.}
\end{figure*}

Filamentous \emph{fd} viruses are monodisperse semi-rigid filaments with \SI{880}{\nm} length, \SI{7}{\nm} diameter, and \SI{2.8}{\um} persistence length~\cite{Barry:2009uv}. When suspended in an aqueous solution at increasing concentrations, they undergo a transition to an aligned nematic phase characterized by long-range orientational order. This isotropic-to-nematic phase transition is quantitatively described by Onsager's theory, indicating that viruses repel one another via hard-core and electrostatic interactions~\cite{Barry:2009uv,Purdy:2003bj,Onsager:1949wv}. Filamentous viruses are chiral and form a twisted nematic (cholesteric) phase in which the director field rotates with a well-defined handedness~\cite{Dogic:2000tp}. For wildtype \emph{fd} virus, the strength of cholesteric interactions is temperature-dependent and continuously increases with decreasing temperature. A single amino acid substitution in the major coat protein leads to the Y21M virus whose cholesteric phase has a handedness opposite to that of the wildtype~\cite{Barry:2009uv}. Mixing wildtype and Y21M viruses produces cholesteric phases with intermediate twist pitches; at a certain ratio, the mixture exhibits no macroscopic twist.

The addition of a non-adsorbing polymer, such as dextran, to a dilute isotropic \emph{fd} suspension induces virus-virus attraction via depletion~\cite{Asakura:1954jy,Asakura:1958kc}. The geometry of the constituent rods ensures that attractive interactions are strongest for lateral associations, causing the viruses to coalesce into one-rod-length-thick, disk-shaped mesoscopic clusters~\cite{Barry:2010cz}. They slowly sediment to the bottom of the glass container, which is coated with a polyacrylamide brush penetrable to dextran in order to suppress depletion-induced virus-wall attractions~\cite{Lau:2009gw}. Over a certain range of depletant concentrations, protrusion fluctuations induce vertical repulsion between clusters, suppressing their face-on association~\cite{Yang:2012ds}. Consequently, such clusters continue to associate laterally, forming large equilibrium 2D colloidal membranes that can be millimeters in diameter [Fig.~\ref{fig:images}(b)]. Single molecule tracking indicates liquid-like order within a membrane. Twisting of constituent chiral viruses is inherently incompatible with assembly into a layered membrane-like structure~\cite{Barry:2008vo}. Consequently, twist can only penetrate into the membrane from the edges and is expelled from the bulk. Unique properties of the colloidal membrane allow for direct visualization of the twist field and quantitative measurement of the twist penetration length $l_\textrm{twist}$~\cite{Barry:2008vo}. When the membrane radius is much bigger than $l_\textrm{twist} \sim \SI{1}{\um}$, the edge adopts a surface-area-minimizing rounded shape with the constituent rods significantly tilting into the membrane plane [Fig.~\ref{fig:images}(b)]; when the membrane radius becomes of the order of $l_\textrm{twist}$ or smaller, the edge profile becomes more square-like and rods do not significantly tilt away from the membrane normal [Fig.~\ref{fig:images}(c)]. Due to thermal excitations, membrane edges undergo ripple fluctuations that can be visualized and precisely quantified [Fig.~\ref{fig:images}(d)].

When chirality-inverted Y21M viruses are used instead of wildtype \textrm{fd}, rods at the edge twist with the opposite handedness, and when the macroscopically achiral mixture of wildtype and Y21M viruses is used, edge-bound rods in each membrane have equal probability of twisting with one handedness or the other~\cite{Gibaud:2012cf}. The achiral mixture exhibits spontaneous symmetry breaking, which has been observed in Langmuir-Blodgett films~\cite{Viswanathan:1994bd,Pettey:1999jx}, another class of two-dimensional structures with nanoscale components, and which has been used in sensors of molecular chirality~\cite{Ohzono:2014kt}.  Increasing the rod chirality raises the free energy of interior untwisted rods while lowering the free energy of edge-bound twisted rods, leading to chiral control of edge line tension~\cite{Gibaud:2012cf}. At sufficiently high chirality, the edge tension approaches zero, and a flat 2D disk spontaneously transitions into an array of 1D twisted ribbons, called a ``starfish''~[Fig.~\ref{fig:images}(e)].  

The twist associated with the membrane’s edge also leads to unconventional pathways of membrane coalescence~\cite{Zakhary:2014de}. As two membranes of same chirality approach each other laterally, the proximal membrane edges can partically coalesce and localize \SI{180}{\degree} of twist to a 1D structure between the membranes; consequently, such structures are called $\pi$-walls~[Fig.~\ref{fig:images}(f)]. The rods twist by \SI{180}{\degree} along the axis connecting the two membranes, from one side of the $\pi$-wall to the other. At the middle of the $\pi$-wall, the rods point in the plane of the membranes.

\section{\label{sec:res}Results}

\subsection{Circular membranes}

In our model, we treat the membrane as a continuous fluid composed of rods at constant density. Once the membrane is stably formed, we assume it does not exchange rods with the surrounding solution; thus, its volume is fixed. The membrane structure is characterized by two coarse-grained degrees of freedom available to the rods: a twist angle $\theta(\ve x)$ about an axis in the membrane plane and a root-mean-square amplitude $b(\ve x)$ of height fluctuations perpendicular to the membrane plane. Perpendicular fluctuations increase the effective thickness of the membrane, and instead of using $b(\ve x)$ directly, we will develop a microscopic theoretical model and present its results using the coarse-grained membrane half-thickness $h(\ve x) = t \cos\theta(\ve x) + b(\ve x)$, where $t$ is the half-length of the virus.

\begin{figure}
	\includegraphics[width=\columnwidth]{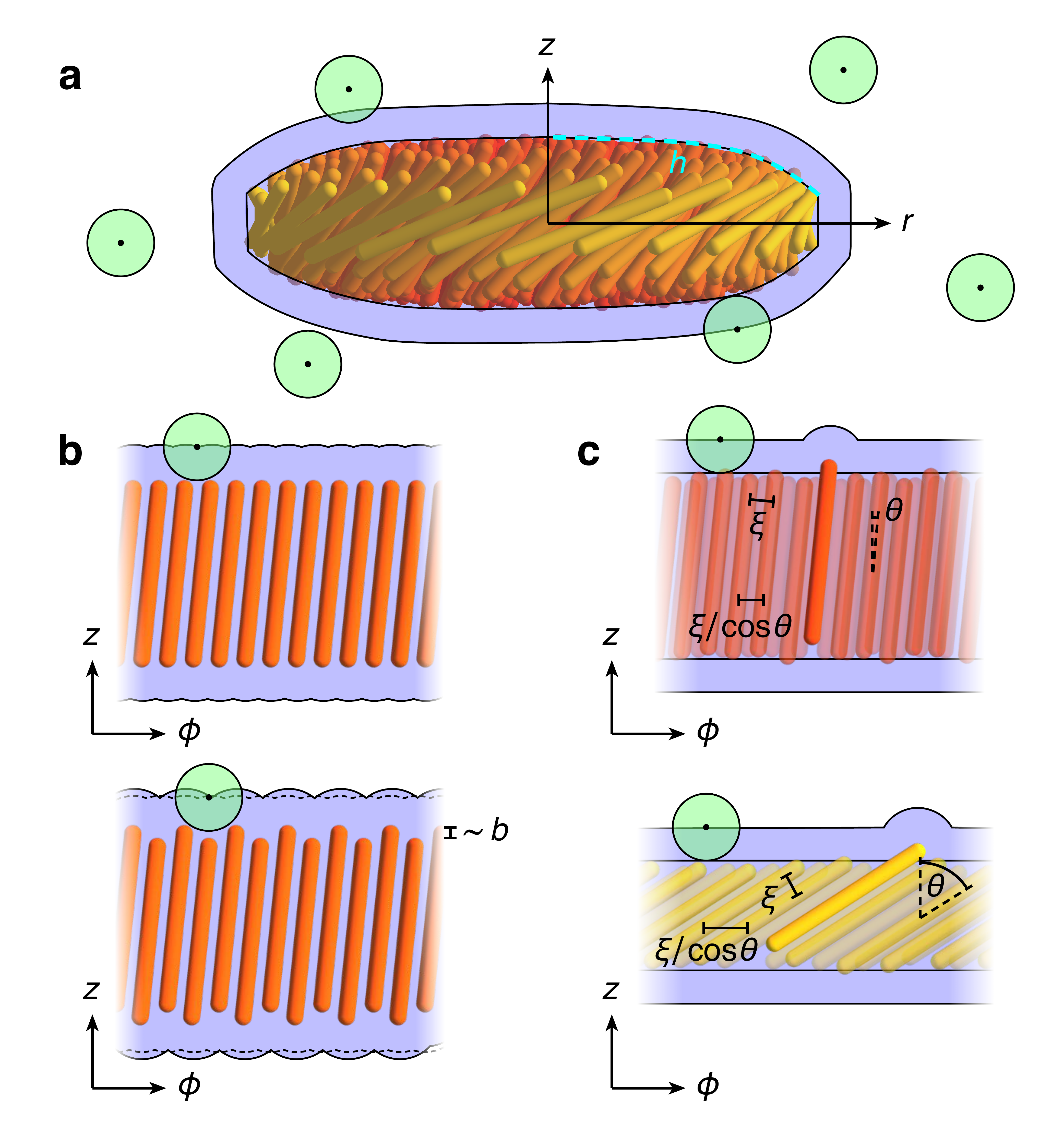}
	\caption{\label{fig:depletion}(Color) Depletant contributions to the membrane free energy. Depleting molecules (dextran polymers in our system) are illustrated in green, and the volume excluded to them by the membrane is illustrated in blue. (a) The volume that is excluded to the depleting polymer due to a  a smooth membrane is comprised of the volume of the membrane itself and to a first order the membrane surface area times the depletant radius. $h(\ve x)$ is the coarse-grained half-thickness of the membrane. (b) Rod height fluctuations produce an effective surface roughness that increases the effective surface area of the membrane and contributes extra excluded volume: (top) an idealized configuration with no height fluctuations and (bottom) an idealized configuration with small-wavelength height fluctuations of amplitude $b = h - t\cos\theta$ that increase the excluded volume. The dotted line represents the excluded volume of the configuration without height fluctuations. (c) The free energy density of rod height fluctuations are calculated in the mean-field limit by considering a single protruding rod amid a membrane of constant local thickness: (top) rods at small tilt angle $\theta$ and (bottom) rods at large $\theta$. The magnitude of this free energy density decreases with increasing $\theta$ because tilted rods are less dense in the membrane plane by a factor of $\cos\theta$, assuming a constant perpendicular distance $\xi$ between rods. In other words, the surface roughness lengthscale in the $\phi$-direction is proportional to $1/\cos\theta$, or equivalently, the extra effective surface area created by fluctuations is proportional to $\cos\theta$. Theoretical details are given in Sec.~\ref{ssec:memb}.}
\end{figure}

The model free energy is comprised of three entropic components. The first term is the Frank free energy that disfavors bend elastic distortions of rods within a membrane while favoring local twisting of rods at their naturally-preferred wavenumber $q$~\cite{Frank:1958ey}; it depends predominantly on $\theta$. All experimental results are obtained using wildtype virus suspensions, which favor left-handed twist~\cite{Barry:2009uv}. The second term is associated with free volume accessible to the depleting polymer due to the presence of the membrane; it depends on the thickness profile $h$ [see Fig.~\ref{fig:depletion}(a)]. Aside from the constant volume of the incompressible membrane, the excluded volume is approximately its surface area times the depletant radius; thus, this term acts as an effective surface tension energy. Its magnitude is proportional to the depletant concentration and to the temperature. The third and final term accounts for the entropy associated with rods protruding from membranes into the surrounding volume occupied by the depleting polymer, a phenomenon reported in Ref.~\cite{Yang:2012ds}. Protrusion of each rod increases the effective surface area of the membrane, which decreases the volume accessible to the depletant molecules. The preferred magnitude of rod height fluctuations $b_0$ is determined by a trade-off between rod entropy, which prefers larger $b$, and the depletion effect, which tends to minimize $b$ [see Fig.~\ref{fig:depletion}(b)]. In our system, the preferred magnitude of this effective surface roughness is very small---$b_0 \ll t$---but the energetic cost of deviations from this value depends on the rod angle $\theta$ [see Fig.~\ref{fig:depletion}(c)]. When $\theta \approx 0$, rods are packed more closely in the plane of the membrane, assuming a constant perpendicular distance $\xi$ between rods. Thus, rod fluctuations produce surface roughness on a smaller length scale, which creates more effective surface area and costs more energy. In this case, $b = b_0$ is strongly preferred, so $h \approx t\cos\theta$ and rod entropy can be ignored. When $\theta \sim 1$, rods are spaced farther apart in the plane of the membrane, leading to fluctuation-produced surface roughness on larger length scales. These longer-wavelength fluctuations resist $b = b_0$ more weakly, so $h$ may differ significantly from $t\cos\theta$. In a similar fashion, manipulating the surface roughness of larger colloids can tune their depletion-induced interaction~\cite{Zhao:2008ij,Kraft:2012cc}. In summary, the rod fluctuation term couples $h$ to $t\cos\theta$ with a $\theta$-dependent coupling strength. To obtain the membrane structure, we minimize the total free energy over $\theta(\ve x)$ and $h(\ve x)$. At the center of the membrane, the membrane is fixed to be one-virus-length thick, while there are no height constraints at the membrane edge.

\begin{table*}
	\caption{\label{tab:1}Membrane parameters and their values.}
	\begin{ruledtabular}\begin{tabular}{ccccc}
		Parameter & Variable & Experimental value & Reference(s) & Theoretical fit value \\
		\hline
		Virus half-length & $t$ & \SI{440}{\nm} & \cite{Barry:2009uv} & same \\
		Temperature & $T$ & 0--\SI{60}{\celsius} & experimental & same \\
		Depletant concentration & $n$ & 35--\SI{51}{\mg\per\mL} & experimental & same \\
		Depletant radius & $a$ & ${\sim}\SI{25}{\nm}$ & \cite{Ioan:2000kl,Armstrong:2004kh,Banks:2005cc}\footnotemark[1] & \SI{31}{\nm} \\
		Nearest-neighbor virus distance & $\xi$ & \SI{12}{\nm} & unpublished\footnotemark[2] & same \\
		Frank elastic constant & $K$ & \SI{0.5}{\pico\N} & \cite{Dogic:2000tp}\footnotemark[3] & \SI{2.8}{\pico\N} \\
		Preferred twist wavenumber & $q(T)$ & $\SI{0.5}{\per\um} \sqrt{1-T/\SI{60}{\celsius}}$ & \cite{Gibaud:2012cf}\footnotemark[3] & $\SI{2.5}{\per\um} \sqrt{1-T/\SI{120}{\celsius}}$  \\
		Virus birefringence & $\Delta n$ & $0.0087\pm0.0007$ & \cite{Barry:2008vo}\footnotemark[4] & $0.0065$
	\end{tabular}\end{ruledtabular}
	\footnotetext[1]{Hydrodynamic radii for dilute solutions of \SI{500}{\kilo\dalton} dextran, whereas our experiments are in the semidilute regime.}
	\footnotetext[2]{Unpublished data extracted from X-ray scattering.}
	\footnotetext[3]{Measured in the bulk cholesteric phase with \emph{fd} virus concentration \SI{100}{\mg\per\mL}, which is lower than the membrane virus concentration \SI{230}{\mg\per\mL} estimated from the experimentally-measured nearest-neighbor virus distance $\xi$.}
	\footnotetext[4]{Assuming that the nematic order parameter in membrane is 1. Membrane virus concentration \SI{230}{\mg\per\mL} estimated from the experimentally-measured nearest-neighbor virus distance $\xi$.}
\end{table*}

In order to obtain quantitatively meaningful results, we use parameter values that are extracted from relevant experimental measurements when possible (Table~\ref{tab:1}). Five parameters, whose values are neither experimentally controlled nor directly measured, are allowed to vary as fit parameters: the characteristic depletant size $a$, the Frank elastic constant $K$ in the one-constant approximation, the amplitude and transition temperature of the temperature-dependent twist wavenumber $q(T)$, and the virus birefringence $\Delta n$. In our theory, we maintain the experimentally-measured square-root behavior of $q(T)$ (see Supplementary Material of~\cite{Gibaud:2012cf}). The Frank elastic constant can be written in dimensionless form as $k(T) \equiv K/natT$, a ratio between the influence of Frank elasticity and that of depletion. Presumably, $K$ depends on temperature in a complicated fashion, as measured for a variety of lyotropic and thermotropic liquid crystals~\cite{DuPre:1975el,Leenhouts:1976hy,Karat:1977ih,Zhou:2012bu}, but we ignore this effect.

\begin{figure*}
	\includegraphics[width=\textwidth]{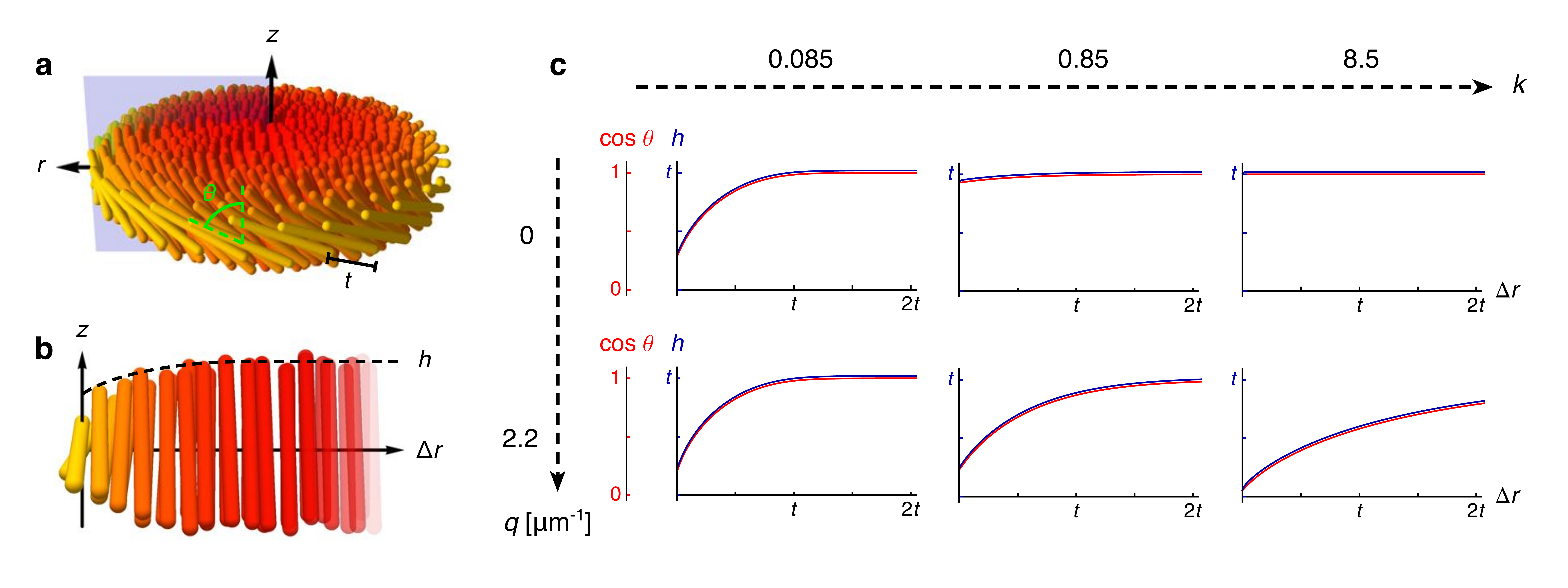}
	\caption{\label{fig:edges}(Color) Vertical edge profile of a large membrane and its dependence on Frank elasticity and chirality. (a) Perspective and (b) cross-section schematics show parametrization of the membrane edge profile and the cylindrical coordinate system. (b) shows rods that intersect the light blue plane in (a). $\Delta r$ is a reverse radial coordinate where $\Delta r = 0$ corresponds to the membrane edge. $h$ is the membrane half-thickness and $\theta$ is the rod tilt angle. $t$ is the half-length of the rods. (c) Calculated edge profiles of a large membrane (radius $R \gg t$) with various Frank-to-depletion ratios $k$ from left to right and preferred twist wavenumbers $q$ from top to bottom. In all cases, $h$ (blue) is almost indistinguishable from $t \cos\theta$ ($\cos\theta$ in red). Note that for $q = 0$ and $k < 1$, $\theta \neq 0$, demonstrating spontaneous symmetry breaking into a configuration with one handedness ($\theta > 0$) or the other ($\theta < 0$). For $q = 0$ and $k \geq 1$, the untwisted state with $\theta = 0$ has lowest energy. Experimental conditions listed in Table~\ref{tab:1} are closest to $k = 0.85$ and $q = \SI{2.2}{\per\um}$.}
\end{figure*}

We first use our theoretical model to determine how membrane structure depends on its radius. We use cylindrical coordinates and assume circular symmetry [Figs.~\ref{fig:edges}(a) and \ref{fig:edges}(b)]. For convenience, we use the reverse radial coordinate $\Delta r$, which originates at the membrane's edge and takes positive values towards the center of the membrane. $\theta$ is the twist angle about the local radial axis. Figure~\ref{fig:edges}(c) plots the vertical membrane profile for membranes with very large radii and varying Frank-to-depletion ratios $k$ and twist wavenumbers $q$. For all conditions, $h \approx t\cos\theta$, indicating that $\theta$ is sufficiently small to suppress rod height fluctuations. Thus, rod entropy does not contribute significantly to the structure of the membrane's edge. First, consider the $q=0$ profiles in Fig.~\ref{fig:edges}(c) corresponding to a macroscopically achiral rod mixture. When $k$ is greater than a critical value $k_c = 1$, the untwisted configuration with $\theta = 0$ is favored. When $k < k_c$, depletion drives spontaneous chiral symmetry breaking into a twisted configuration with either $\theta > 0$ or $\theta < 0$. In the $k \rightarrow 0$ limit where only depletion exists, the vertical edge profile is semicircular to minimize the membrane surface area. Now, consider the $q = \SI{2.2}{\per\um}$ case in Fig.~\ref{fig:edges}(c) corresponding to a chiral rod mixture. Twisted configurations of one handedness (here, $\theta > 0$ for $q > 0$) become favored at all $k$. In the depletion-dominated regime $k \ll 1$, the vertical edge profile again approaches a semicircle. In the Frank-elasticity-dominated regime $k \gg 1$, the rod twist decays with penetration length $l_\textrm{twist} \approx \sqrt{k}/t$, in analogy to the way that twist penetrates into a smectic phase. Calculations of $k_c$ and $l_\textrm{twist}$ are provided in Appendix~\ref{sec:breaking}.
\begin{figure*}
	\includegraphics[width=\textwidth]{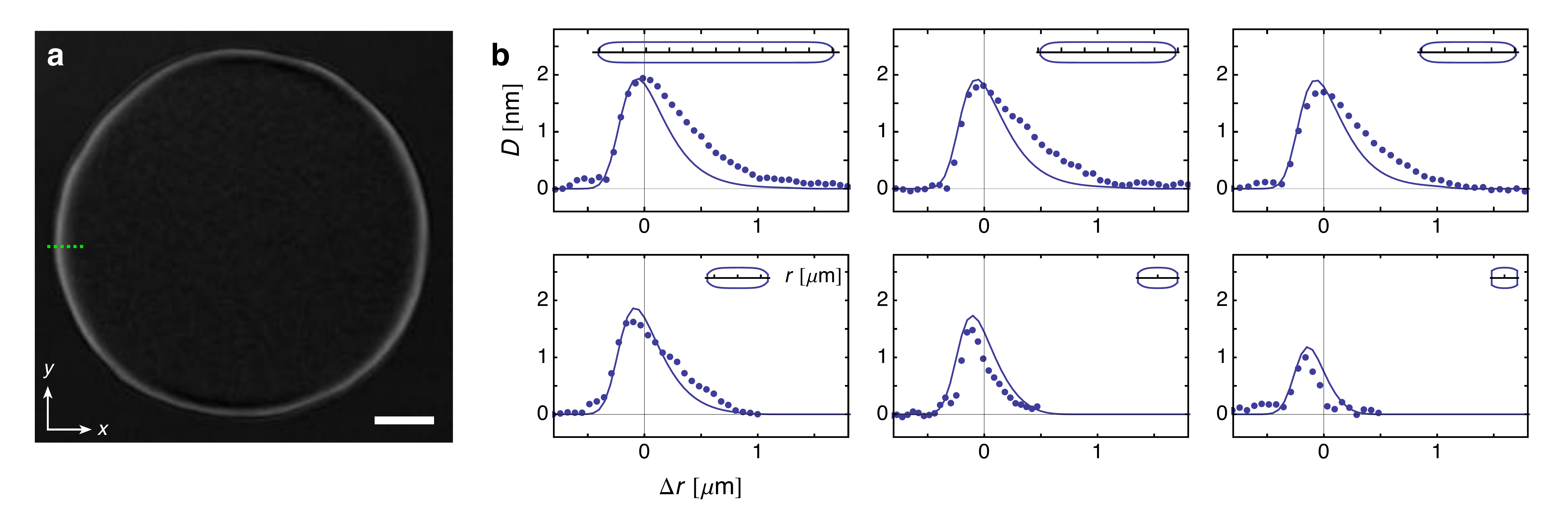}
	\caption{\label{fig:edgeret}(Color) Membrane edge retardance. (a) 2D LC-PolScope birefringence map of a large circular membrane with retardance represented as pixel brightness. The dotted green line approximately corresponds to the range of $\Delta r$'s plotted in (b). Scale bar, \SI{4}{\um}. (b) Retardance values $D$ for circular membranes of various sizes. The points indicate experimental data at temperature $T = \SI{22}{\celsius}$ and depletant concentration $n = \SI{45}{\mg\per\mL}$. The lines indicate theoretical results calculated with these parameter values and those described in Table~\ref{tab:1}, giving $k = 0.85$ and $q = \SI{2.2}{\per\um}$. Membrane radii range from $\SI{5.1}{\um}$ (top left) to $\SI{0.45}{\um}$ (lower right). The insets show the calculated membrane profile $h(r)$, plotted with an aspect ratio of 1. Tick marks signify \SI{1}{\um} increments. $t \cos \theta$, not shown, is strongly coupled to $h$ in all cases.}
\end{figure*}

In addition to describing edges of large membranes, our theoretical model also describes how edge profile varies with decreasing membrane diameter. To test these predictions, we measure the retardance of different-sized membranes using quantitative LC-PolScope microscopy, which directly reveals the twisting of rods away from the membrane normal. When polarized light passes through a birefringent material, the components corresponding to the dielectric tensor eigenvectors---the ordinary and extraodinary waves---propagate at different speeds. The resulting phase difference between these components multiplied by the wavelength of the light is the retardance $D$. For a uniaxial crystal of constant thickness, retardance can be calculated as $D = \Delta n h \sin^2\theta$~\cite{bornwolf}, where $\Delta n$ is the birefingence. For membranes of various radii, we calculate $D(\Delta r)$ with the fit values given in Table~\ref{tab:1} and the approximation $h = t\cos\theta$, since our results in Fig.~\ref{fig:edges}(c) demonstrate that rod fluctuations $b$ are insignificant for membrane edges. We use the same parameter values for all membrane sizes; only the radius changes. The radially-averaged edge retardance profiles measured for membranes of various radii match well with our theoretical predictions [Fig.~\ref{fig:edgeret}(b)]. These results demonstrate that rods are less tilted at the edges of smaller membranes compared to those of larger membranes [insets of Fig~\ref{fig:edgeret}(b)], consistent with observations that larger membranes appear on side-view to have rounded edges while smaller membranes have squared-off edges [Figs.~\ref{fig:images}(b) and \ref{fig:images}(c)].

\begin{figure}
	\includegraphics[width=\columnwidth]{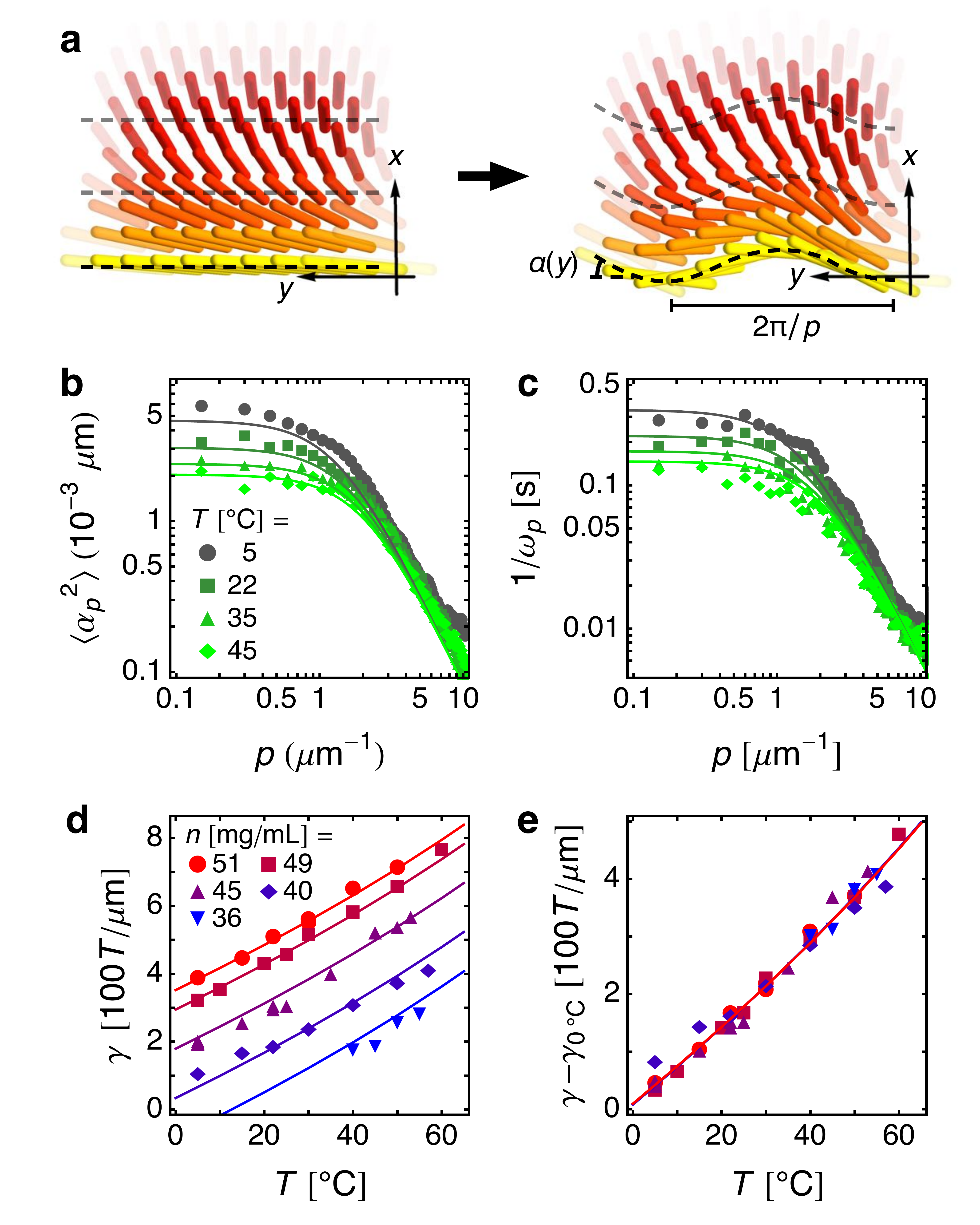}
	\caption{\label{fig:tension}(Color) Line tension analysis of the membrane edge. The points indicate experimental data at various temperatures $T$ and depletant concentrations $n$. The lines indicate theoretical results calculated for corresponding parameter values and those described in Table~\ref{tab:1}. (a) Schematic of the membrane ripple ansatz through which line tension and line bending modulus are calculated. $\alpha(y)$ is the angle between the ripple tangent vector and the $y$-axis. (b) Thermal fluctuation amplitudes $\avg{\alpha_p^2}$ and (c) autocorrelation decay timescales $1/\omega_p$ of ripple fluctuations for depleting concentration $n = \SI{45}{\mg\per\mL}$ and various temperatures $T$. The theoretical plots of $1/\omega_p$ use the fit value for the 1D membrane edge viscosity $\eta_\textrm{1D} = \SI{300}{\milli\pascal\s\square\um}$. (d) Line tension $\gamma$ and (e) its relative temperature-dependent behavior as a function of temperature for various $n$. For each $n$, $\gamma_{\SI{0}{\celsius}}$ is the line tension extrapolated to $T = \SI{0}{\celsius}$.}
\end{figure}

With detailed understanding of the membrane's edge structure, we next study its fluctuations, which are clearly visible and easily quantified with optical microscopy [Fig.~\ref{fig:images}(d)]. In the large membrane limit, we ignore curvature of the edge and, with Cartesian coordinates, place the very edge at $x=0$ [Fig.~\ref{fig:tension}(a)]. $\theta$ is now the twist angle about the $x$-axis. Using the previously discussed model, we first calculate $h(x)$ and $\theta(x)$ for a flat edge. We then introduce a small ripple at the edge with the tangent angle $\alpha(y)$ that perturbs the rod configuration as in Fig.~\ref{fig:tension}(a). The unperturbed configuration along lines parallel to the $y$-axis is mapped onto curves with the same tangent angle $\alpha(y)$, and the rod rotation axis for $\theta$ is always perpendicular to these curves. See Section~\ref{ssec:ripple} for a mathematical description of this ripple ansatz. We can write $\alpha(y)$ in terms of Fourier components $\alpha_p$, where $p$ is the ripple wavenumber. To lowest order in these Fourier components, the relative free energy per unit length is $f = \frac{1}{2} \sum_p (\gamma[h,\theta] + \kappa[h,\theta]p^2) \alpha_p^2$, where the line tension $\gamma[h,\theta]$ and the edge bending modulus $\kappa[h,\theta]$ are functionals of the flat edge configuration. The line tension describes the energetic cost of having an edge interface, and the edge bending modulus arises from the rod director's bend distortion introduced by the ripple. By equipartition and viscous hydrodynamics, we obtain the fluctuation spectra
\begin{equation}
	\avg{\alpha_p^2} = \frac{\kT}{\gamma + \kappa p^2}, \quad \frac{1}{\omega_p} = \frac{\eta_\textrm{1D}}{\gamma + \kappa p^2}.\nonumber
\end{equation}
$\langle \alpha_p^2 \rangle$ is the average fluctuation amplitude of Fourier mode $p$ and $\omega_p$ is the temporal autocorrelation decay constant of Fourier mode $p$, as found in the temporal autocorrelation function $\langle \alpha_p(t)\alpha_p(0) \rangle = \langle \alpha_p^2 \rangle \ee^{-\omega_p t}$. The decay of fluctuation correlations arises from dissipative forces, the most significant of which are membrane viscous stresses since we expect the membrane to be much more viscous than the solvent. $\eta_\textrm{1D}$ is the one-dimensional (1D) viscosity of the membrane edge.

Using the fit values in Table~\ref{tab:1} describing the membrane edge, our theoretical model predicts values for $\gamma$ and $\kappa$, which determine the fluctuation spectra $\langle \alpha_p^2 \rangle$ and $1/\omega_p$. These predictions can be tested experimentally, and the value of $\gamma$ can be extracted from the low-$p$ limit of $\langle \alpha_p^2 \rangle$. The experimental and theoretical spectra match well over a variety of temperatures [Figs.~\ref{fig:tension}(b) and \ref{fig:tension}(c)]. These calculations still assume $h = t\cos\theta$, since Fig.~\ref{fig:edges}(c) demonstrates that rod fluctuations $b$ are insignificant for membrane edges. The ratio between $1/\omega_p$ and $\langle \alpha_p^2 \rangle$ appears constant for all measured values of $p$---in agreement with our theory---and gives a value for $\eta_\textrm{1D} \approx \SI{300}{\milli\pascal\s\square\um}$. 

We expect the 3D membrane viscosity $\eta$ to be strongly inhomogeneous and anisotropic at the edge due to the large aspect ratio of the rods. For instance, during a ripple fluctuation, rods oriented more vertically may slide past each other more easily than those tilted more horizontally. To roughly estimate the magnitude of $\eta$, we write $\eta_\textrm{1D} \sim \int \dd x\,\dd z\,\eta \sim A \eta$, where $A \sim 2 t l_\textrm{twist} \sim 2 t^2$ is an estimated cross-sectional area of the membrane edge participating in these ripple fluctuations. As calculated in Appendix~\ref{sec:breaking}, $l_\textrm{twist} \approx \sqrt{k}t$ is the twist penetration depth, and the parameter values provided in Table~\ref{tab:1} satisfy $k \sim 1$. This gives $\eta \sim \SI{800}{\milli\Pa\s}$, much greater than the solvent viscosity, which is $\eta_\textrm{s} \approx \SI{3}{\milli\Pa\s}$ for 5 w\% \SI{500}{\kilo\dalton} aqueous dextran~\cite{Cush:1997ky}.

Measurements and calculations of the line tension $\gamma$ show good quantitative agreement over a variety of temperatures $T$ and depletant concentrations $n$ [Fig.~\ref{fig:tension}(d)]. For all $n$, $\gamma$ decreases as $T$ is reduced. If we measure $\gamma$ relative to its value at a standard temperature, say $T = \SI{0}{\celsius}$, the line tensions for different $n$ all collapse onto a single curve [Fig.~\ref{fig:tension}(e)], indicating that the relative effect of temperature change on $\gamma$ is independent of the depletant concentration. In Sec.~\ref{ssec:ripple}, we see how these effects arise naturally in our model via a $q$-dependent chiral term in the line tension. Colloidal membranes assembled from chiral rods are inherently frustrated, because the particles cannot simultaneously twist locally and assemble into a monolayer globally. Consequently, twist is expelled from the membrane interior and localized to its edges. Note that $q(T)$ is a monotonically decreasing function of $T$ (Table~\ref{tab:1}). Decreasing the temperature increases $q(T)$ and lowers the free energy of edge-bound twisted rods, leading to chiral control of edge line tension~\cite{Gibaud:2012cf}.

\subsection{Starfish morphological transition and $\pi$-wall structure}

\begin{figure}
	\includegraphics[width=\columnwidth]{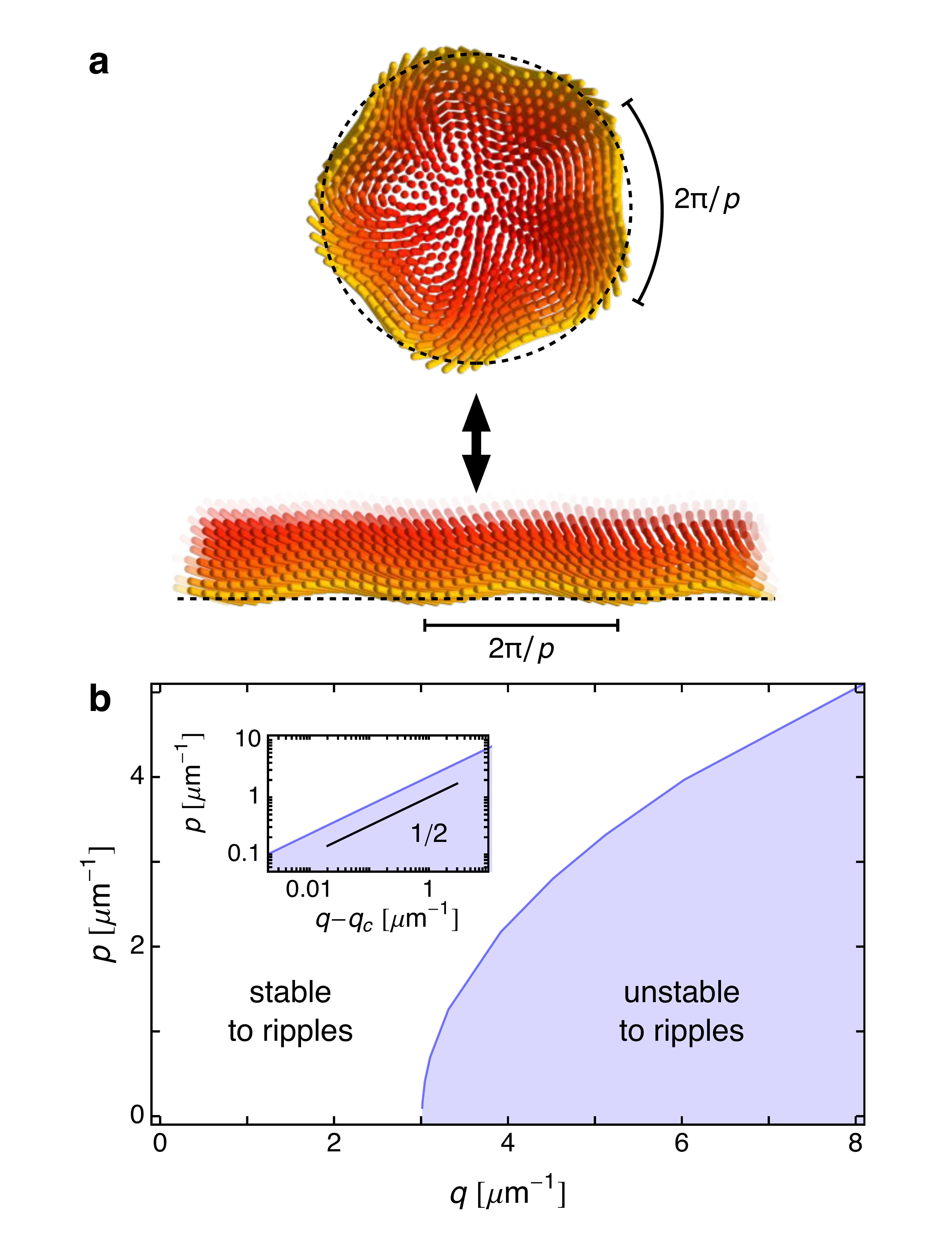}
	\caption{\label{fig:starfish}(Color) Starfish instability transition for large membranes. (a) Starfish arms grow from unstable ripple fluctuations in our theory. For large membranes, we can take the Cartesian limit and ignore the curvature of the edge. (b) The shaded region indicates unstable ripple wavenumbers $p$ calculated for preferred twist wavenumbers $q$ and constant Frank-to-depletion ratio $k = 0.85$. We take $p$ to be continuous, corresponding to the infinite membrane size limit; for finite-sized circular membranes, continuity permits only certain values of $p$, namely multiples of the inverse radius. The inset plots the same results on a log-log scale to demonstrate that as $q$ increases past a critical $q_c \approx \SI{3}{\per\um}$, the range of unstable $p$'s grows as a power law with exponent 1/2.}
\end{figure}

We now apply our theory to explain more exotic structures found in colloidal membranes. For example, when circular membranes are subjected to a temperature quench, the line tension decreases significantly and fluctuations at the edge increase in amplitude. For sufficiently low $T$, the circular membrane becomes unstable and grows arms of twisted ribbons along its entire periphery [Fig.~\ref{fig:images}(e)]. In our model, these starfish arms arise from the aforementioned ripple fluctuations [Fig.~\ref{fig:starfish}(a)]. As the temperature decreases, the chiral wavenumber $q(T)$ increases and lowers the line tension $\gamma$. For sufficiently large $q$, $\gamma$ becomes negative and long-wavelength ripple modes along the membrane circumference become unstable, which presumably grow and twist into starfish arms. Figure~\ref{fig:starfish}(b) plots the range of unstable wavenumbers $p$, measured around the circumference, as a function of chiral wavenumber $q$ for constant $k = 0.85$. Above a critical $q_c \approx \SI{3}{\per\um}$, low-$p$ modes become unstable. An instability with $p \approx \SI{1}{\per\um}$ in a membrane of radius $R \approx \SI{5}{\um}$ corresponds to a five-armed starfish structure as depicted in Figs.~\ref{fig:images}(e) and \ref{fig:starfish}(a), so the order of magnitude of unstable $p$'s calculated in Fig.~\ref{fig:starfish}(b) follows expectations. Note that changing the temperature also changes $k(T)$, but the effect is qualitatively insignificant. The transition from a circular membrane to a starfish structure is reversible, so reheating to a positive $\gamma$ drives the edge-length-maximizing starfish structure to decrease its edge length and become circular again~\cite{Gibaud:2012cf}.

\begin{figure*}
	\includegraphics[width=\textwidth]{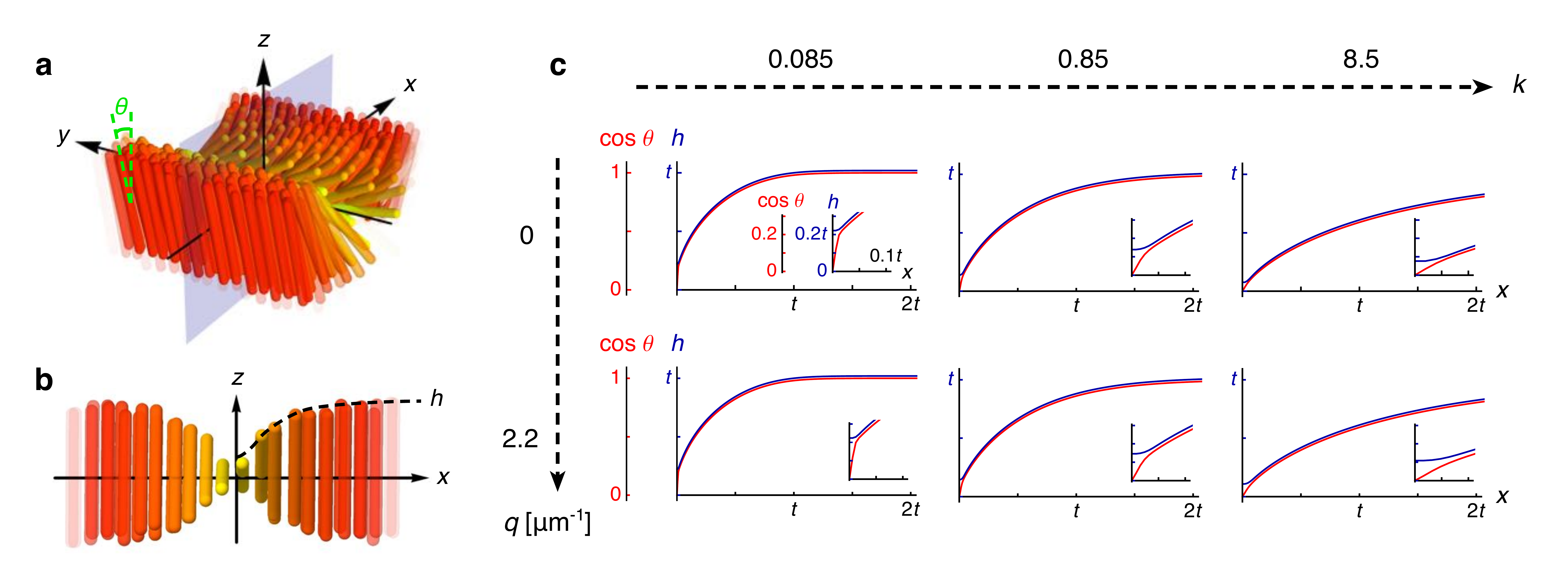}
	\caption{\label{fig:walls}(Color) Vertical $\pi$-wall profiles and their dependence on Frank elasticity and chirality. (a) Perspective and (b) cross-section schematics showing parametrization of $\pi$-wall profile and Cartesian coordinate system. (b) shows rods that intersect the light blue plane in (a). $h$ is the membrane half-thickness and $\theta$ is the rod tilt angle. $t$ is the half-length of the rods. (c) Calculated vertical $\pi$-wall profiles for various Frank-to-depletion ratios $k$ from left to right and preferred twist wavenumbers $q$ from top to bottom. In all cases, $h$ (blue) is almost indistinguishable from $t \cos\theta$ ($\cos\theta$ in red) away from $x = 0$. Near $x = 0$, $h$ approaches a finite mid-wall value while $\cos\theta$ approaches 0. Insets highlight the profile near $x = 0$. Experimental conditions listed in Table~\ref{tab:1} are closest to $k = 0.85$ and $q = \SI{2.2}{\per\um}$.}
\end{figure*}

\begin{figure}
	\includegraphics[width=\columnwidth]{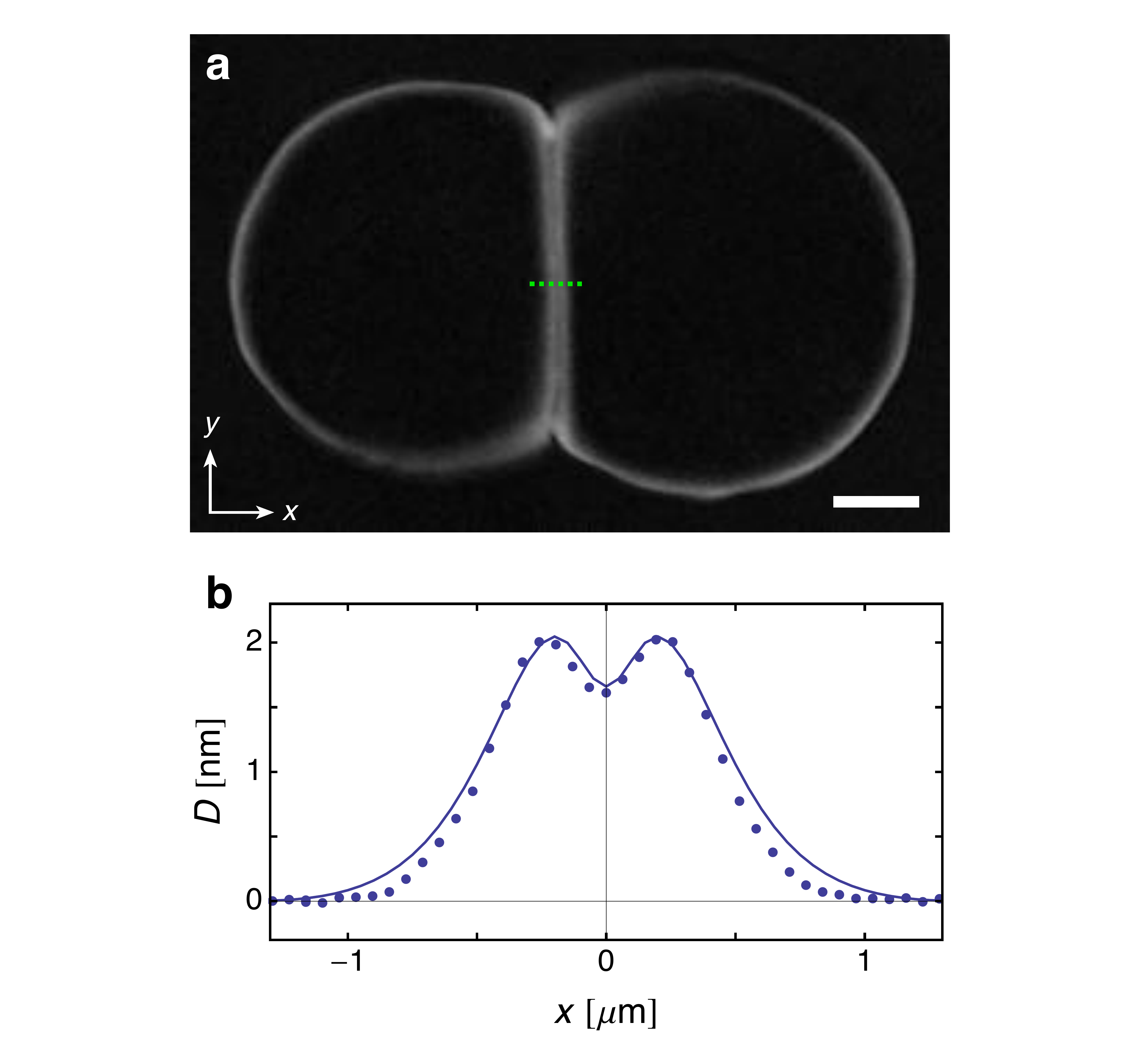}
	\caption{\label{fig:wallret}(Color) $\pi$-wall retardance. (a) 2D LC-PolScope birefringence map of two circular membranes joined through a $\pi$-wall with retardance represented as pixel brightness. The dotted green line approximately corresponds to the range of $x$'s plotted in (b). Scale bar, \SI{4}{\um}. (b) Retardance values $D$. The points indicate experimental data at temperature $T = \SI{22}{\celsius}$ and depletant concentration $n = \SI{45}{\mg\per\mL}$. The lines indicate theoretical results calculated with these parameter values and those described in Table~\ref{tab:1}.}
\end{figure}

We also use our theoretical model to quantitatively explain another prominent and experimentally-characterized feature of colloidal membranes: the $\pi$-wall. Observations on the assembly pathways and structure of $\pi$-walls were described in Sec.~\ref{sec:over} and Fig.~\ref{fig:images}(f). Briefly, two membranes of the same handedness can partially coalesce into a single membrane and trap a twist domain wall, or $\pi$-wall, through which the rod director twists by \SI{180}{\degree}. To investigate these structures theoretically, we use Cartesian coordinates [Figs.~\ref{fig:walls}(a) and \ref{fig:walls}(b)] and set $\theta(x=0) = \pi/2$ so the rods at the middle of the $\pi$-wall lie completely in the membrane plane. Fig.~\ref{fig:walls}(c) plots the thickness profile over a range of Frank-to-depletion ratios $k$ and chiral twist wavenumbers $q$. In all cases, $h$ is much greater than $t\cos\theta$ at the middle of the wall, since the coupling that sets $h \approx t\cos\theta$ becomes very weak when $\theta\approx\pi/2$. Remember, $h = t\cos\theta + b$, where $b$ is the amplitude of rod fluctuations perpendicular to the membrane. This means these rods undergo position fluctuations in the $z$-direction that are many times larger than both their projected height $t\cos\theta$ and their diameter $d \approx 0.02t$. Such a phenomenon would require rods to pass through each other, which is theoretically allowed because we ignore rod-rod interactions, but we wish to interpret this result physically. $h \gg t\cos\theta$ and $h \gg d$ indicate that the membrane is thicker than multiple layers of tilted rods, so these large fluctuations may be physically manifested as rods stacking on top of each other. The addition of repulsive rod-rod interactions may further increase the thickness of the $\pi$-wall. As for the $\pi$-wall profile, similarly to the membrane edge, the depletion-dominated regime $k \ll 1$ leads to a circular profile and the Frank-elasticity-dominated regime $k \gg 1$ leads to slow rod twist decay [Fig.~\ref{fig:walls}(c)]. $q$ does not significantly affect the $\pi$-wall profile among the parameter values explored; indeed, it appears in a $\theta$-dependent free energy term that can almost be integrated to the boundary, and $\theta(x=0)$ and $\theta(x\rightarrow \infty)$ are fixed (see Section~\ref{ssec:memb}). With the same parameters used to calculate the edge structure (Table~\ref{tab:1}), the calculated optical retardance of the $\pi$-wall quantitatively matches the experimentally measured profile (Fig.~\ref{fig:wallret}).

\section{\label{sec:th}Theoretical development}

\subsection{\label{ssec:memb}Membrane parametrization and free energy}

We treat the membrane as a continuous medium composed of rods at constant density, and we fix the number of rods in the membrane by fixing the membrane volume. The coarse-grained rod twist angle $\theta(\ve x)$, rod height fluctuation amplitude $b(\ve x)$, and membrane half-thickness are related by $h(\ve x) = t \cos\theta(\ve x) + b(\ve x)$, where $t$ is the half-length of the virus. We will first develop the model assuming a circularly-symmetric membrane of radius $R$ and using cylindrical coordinates in which $h(r)$, $b(r)$, and $\theta(r)$ only depend on the radial coordinate.

We model the rods as liquid crystals whose orientations are described by a Frank elastic free energy~\cite{Frank:1958ey}. In a circular geometry, the rods point in the $z$-direction but can twist with angle $\theta$ in the azimuthal direction [Figs.~\ref{fig:edges}(a) and \ref{fig:edges}(b)]. Using the one-constant approximation, the free energy is:
\begin{eqnarray}
	F_\textrm{Frank} &=& K \int \dd^2 \ve x\, h\big[(\bdiv\ve n)^2 + (\bcurl\ve n)^2 \nonumber\\
		& & \qquad\qquad\qquad{}- 2 q \ve n\cdot\bcurl\ve n\big]
	\label{eqn:ffrankgen}\\
	&=& 2\pi K \int_0^R \dd r\, h \bigg[r (\partial_r\theta)^2 + \sin 2\theta\, \partial_r\theta + \frac{\sin^2\theta}{r} \nonumber\\
		& & \qquad\qquad\qquad{}- 2qr \partial_r\theta - q \sin2\theta\bigg].
	\label{eqn:ffrank}
\end{eqnarray}
$K$ is the 3D Frank elastic constant and $q$ is the preferred twist wavenumber associated with intrinsic chirality of the constituent rods. $\ve n(r) =\sin\theta(r)\veh \phi + \cos\theta(r)\veh z$ is the nematic director. The $q$ term breaks chiral symmetry, such that for $q > 0$, twisted membranes with $\partial_r\theta > 0$ have lower energy than those with $\partial_r\theta < 0$. When $q = 0$, the total free energy is invariant under the chirality inversion $\theta \rightarrow -\theta$.

The depletant polymers act to minimize the volume excluded to them by the membrane. For polymers small compared to the dimensions of the membrane, this excluded volume is approximately $V_0 + a A$, where $V_0$ is the volume of the membrane, $A$ is the surface area of the membrane, and $a$ is the characteristic depletant radius~\cite{Allendoerfer:1948da} [see Fig.~\ref{fig:depletion}(a)]. $V_0$ is constant, so depletion serves as an effective surface tension. Consequently, the free energy is given by:
\begin{eqnarray}
	F_\textrm{dep} &=& 2na\kT \left[\int \dd^2 \ve x \sqrt{1 + (\bnabla h)^2} + \int \dd l\, h\right]
	\label{eqn:fdepgen}\\
	&=& 4\pi na\kT \left[\int_0^R \dd r\, r \sqrt{1 + (\partial_r h)^2} + Rh(R)\right],
	\label{eqn:fdep}
\end{eqnarray}
where $n$ is the depletant concentration, $T$ is the temperature, and $k_B$ is the Boltzmann constant. $\int \dd l$ indicates an integral over the membrane edge boundary.

Finally, we allow rods to fluctuate perpendicularly to the membrane plane. In general, these fluctuations have complicated, non-linear effects on the free energy, but for simplicity, we only consider fluctuations of single rods and ignore their interactions and correlations~\cite{Yang:2012ds}. When a single rod at small tilt angle $\theta$ protrudes by a small perpendicular distance $z$ above a flat coarse-grained membrane surface, it introduces an additional spherical cap of volume $\pi a z^2$ that is excluded to the depleting polymers [see Fig.~\ref{fig:depletion}(c)]. Meanwhile, these protrusions are entropically favored by the rods. For a distribution of vertical rod displacements $p(z)$, the fluctuation free energy for a single rod is a sum of depletant and rod entropy contributions: 
\begin{equation}
	F_\textrm{single} = \pi na\kT \int \dd z\,p(z)z^2 + \kT\int\dd z\,p(z)\log p(z). \nonumber
\end{equation}
Minimizing this free energy yields $p_0(z) = (2\pi b_0^2)^{-1/2} \exp(-z^2/2b_0^2)$, where $b_0 = (2\pi na)^{-1/2}$.

If all rods were to fluctuate with the preferred amplitude $b_0$, then the membrane half-thickness $h$ and rod angle $\theta$ would be exactly related as $h = t\cos\theta + b_0$. However, in certain structures such as the mid-planes of $\pi$-walls, the Frank and depletion free energies favor profiles $h(x)$ and $\theta(x)$ that significantly deviate from this relationship. To propertly describe these structures and account for the energetic cost of $h \neq t\cos\theta + b_0$, we calculate the free energy of Gaussian rod fluctuations of amplitude $b \neq b_0$. Using the distribution $p(z) = (2\pi b^2)^{-1/2} \exp(-z^2/2b^2)$, the single-rod free energy becomes $F_\textrm{single} = 2\pi na\kT (b-b_0)^2$ to leading order in $b-b_0$. To coarse-grain this expression, we multiply by the rod density and integrate over the membrane area. For simplicity, we assume the rods are packed hexagonally and maintain a constant perpendicular distance $\xi$ between nearest-neighbors. In the small $\theta$ limit, the area occupied by each rod is $\sqrt{3} \xi^2/\cos\theta$. Our final expression for the rod fluctuation free energy is
\begin{equation}
	F_\textrm{rod} = \frac{8\pi^2 na\kT}{\sqrt{3}\xi^2} \int_0^R \dd r\, r \cos\theta \left(h-t\cos\theta-b_0\right)^2,
	\label{eqn:frod}
\end{equation}
where we have written $b$ in terms of $h$ and $\theta$. This term allows $h$ to deviate from $t\cos\theta + b_0$ with an energy penalty corresponding to the magnitude of the deviation. Heuristically, the energy penalty is proportional to $\cos\theta$ because at higher $\theta$, the rods are spaced farther apart in the plane of the membrane, so height fluctuations of individual rods induce less roughness at the membrane surface [Fig.~\ref{fig:depletion}(c)].

We minimize the total free energy with volume-conserving Lagrange multiplier $\lambda$
\begin{equation}
	F = F_\textrm{dep} + F_\textrm{Frank} + F_\textrm{rod} + \lambda \left[V_0 - 4\pi \int_0^R \dd r\, rh\right] 
	\label{eqn:f}
\end{equation}
over $h(r)$ and $\theta(r)$ to obtain the edge profile. The boundary conditions are $h(0) = t + b_0$ and $\theta(0) = 0$; $h(R)$ and $\theta(R)$ are free.

Equation~\ref{eqn:f} simplifies for large membranes when $R$ is much greater than the penetration depth of edge twist $l_\textrm{twist}$; the edge becomes essentially straight. We can then study the profile of a twisted membrane formed from an untwisted rectangular membrane of length $L_y\rightarrow \infty$ along the $y$-direction and length $2L_x \ll L_y$ along the $x$-direction. We allow the membrane profile to vary along the $x$-direction and impose reflection symmetry about the midline $x = L_x$ where the rods are perpendicular to the membrane (analogous to $r = 0$ for the original circular geometry). We are interested in the edge profile at $x = 0$. In this setup, each free energy integral becomes its Cartesian version, with $F_\textrm{Frank}$ losing bend distortion terms that arise from a circular geometry. Instead of a Lagrange multiplier term, however, volume conservation can be directly enforced in the following way. The volume of the half of the untwisted membrane between $x=0$ and $x=L_x$ is $V_0 = 2 (t+b_0) L_x L_y$. The change in volume brought about by a varying $h(x)$ is $\Delta V = 2 L_y \int_0^{L_x}\dd x\,[h(x)-(t+b_0)]$.  To compensate for the lost volume, we introduce extra volume at the membrane midline where $h(x) = t + b_0$ by adding a width $\Delta L_x$ of untwisted rods; volume conservation requires $\Delta L_x = -\Delta V/[2(t+b_0)L_y] = \int_0^{L_x}\dd x\,[1-h(x)/(t+b_0)]$. This extra width increases the half-membrane's surface area by $\Delta A = 2 L_y \Delta L_x$ and, since depletion free energy is proportional to surface area, contributes the additional term $na\kT\Delta A$ to $F_\textrm{dep}$. Ignoring a constant term proportional to $L_x L_y$, the total free energy becomes
%\begin{widetext}
%\begin{equation}
%	\frac{F}{2na\kT L_y} = \int_0^{L_x}\dd x\left(\sqrt{1+(\partial_x h)^2} - \frac{h}{t+b_0}\right) + h(0) + \frac{k}{2} \int_0^{L_x}\dd x\,h\left[(\partial_x\theta)^2 + 2 q \partial_x\theta\right] + \frac{2\pi}{\sqrt{3}\xi^2} \int_0^{L_x}\dd x\cos\theta (h-t\cos\theta-b_0)^2.
%	\label{eqn:fcartesian}
%\end{equation}
%\end{widetext}
\begin{eqnarray}
	\frac{F}{2na\kT L_y} &=& \int_0^{L_x}\dd x\left(\sqrt{1+(\partial_x h)^2} - \frac{h}{t+b_0}\right) + h(0) \nonumber\\
	&& {}+ \frac{kt}{2} \int_0^{L_x}\dd x\,h\left[(\partial_x\theta)^2 + 2 q \partial_x\theta\right] \nonumber\\
	&& {}+ \frac{2\pi}{\sqrt{3}\xi^2} \int_0^{L_x}\dd x\cos\theta (h-t\cos\theta-b_0)^2,
	\label{eqn:fcartesian}
\end{eqnarray}
where again, $k = K/nat\kT$. Strictly speaking, the integrals in the last two terms should extend from $0$ to $L_x + \Delta L_x$, but the contributions to the integrals from $L_x$ to $L_x+\Delta L_x$ are zero because $\partial_x \theta = 0$, $\theta = 0$, and $h=t+b_0$ in the interior of the membrane. Comparing Eqs.~\ref{eqn:f} and \ref{eqn:fcartesian}, the additional surface area term is analogous to a Lagrange multiplier with value $na\kT/(t+b_0)$, the effective osmotic pressure exerted by the depletants on the membrane. Also, since this Cartesian parametrization implicitly inverts the membrane orientation compared to the cylindrical parametrization (instead of decreasing $r$, increasing $x$ moves into the interior of membrane), the $q$-term in Eq.~\ref{eqn:fcartesian} has the opposite sign of the $q$-terms in Eq.~\ref{eqn:ffrank}.

For membrane edges calculated in Fig.~\ref{fig:edges}(c), $h \approx t\cos\theta$, which means rod height fluctuations $b$ are strongly suppressed. This motivates simplification of the free energy by taking the infinite coupling limit in which $F_\textrm{rod}$ enforces $h = t\cos\theta + b_0$ and therefore disappears from the free energy. Using values in Table~\ref{tab:1}, we calculate $b_0 \approx 0.03t$ and make the further approximation that these protrusion fluctuations contribute only a small fraction to the membrane thickness and can thus be neglected: $b_0 = 0$. Numerical calculations of all edge properties fixing $h = t\cos\theta$ are indistinguishable from those using the full theory. Thus, the precise form of $F_\textrm{rod}$, whose derivation required many assumptions, does not matter for membrane edges as long as it strongly couples $h$ to $t\cos\theta$. This simplification permits derivation of some analytical results, including an investigation into spontaneous chiral symmetry breaking for $q = 0$, which are given in Appendix~\ref{sec:breaking}.

For $\pi$-walls, we use Eq.~\ref{eqn:fcartesian} without the boundary depletion term proportional to $h(0)$ because $x = 0$ is the middle of the wall and no longer an edge boundary. The rods there must lie in the membrane plane, so we gain the extra boundary condition $\theta(0) = \pi/2$. Now $\theta$ is fixed at both boundaries, so if $h$ were enforced to be a function of $\theta$ like $h = t\cos\theta$, the $q$-term could be integrated to a constant and the profiles would not depend on $q$. However, unlike their counterparts at edges, $h$ and $\theta$ are independent near $x = 0$, where calculations show that the vertical mid-wall profile satisfies $h \gg t\cos\theta$; thus, the membrane structures depend slightly on $q$ [Fig.~\ref{fig:walls}(c)]. This independence arises due to the angle-dependent coupling strength of $F_\textrm{rod}$, which has a factor of $\cos\theta$ in the integrand (Eq.~\ref{eqn:frod}). Away from the middle of the wall, $\cos\theta \approx 1$ and deviations from $h = t\cos\theta + b_0$ are costly for $F_\textrm{rod}$. As $\cos\theta$ approaches 0, these deviations cost less energy in $F_\textrm{rod}$, so other terms such as $F_\textrm{dep}$ (Eq.~\ref{eqn:fdep} without the boundary term) gain influence on the profile configuration. The competition between $F_\textrm{rod}$, which prefers $h$ to decrease with $\cos\theta$ towards the middle of the wall, and $F_\textrm{dep}$, which prefers a constant $h$, sets the mid-wall thickness.

It is worthwhile at this point to compare our theory with an alternative one, which we will refer to as the KM theory after its developers Kaplan and Meyer~\cite{Zakhary:2014de,Kaplan:2014cz}, that also produces results in very good agreement with experimental observations. First, it should be emphasized that the philosophical approaches of the two theories are different. Ours can be viewed as a minimalist theory based directly on entropic interactions induced by dextran depletants and to a lesser extent by the viruses themselves. The KM theory, in the grand tradition of liquid-crystal physics, is phenomenological at its core. It introduces an order parameter $\Psi$, inspired by that describing order in a 3D smectic, that describes the transition from rods oriented predominantly perpendicular to the membrane plane (``smectic'' phase with $\Psi\neq 0$) to rods oriented predominantly parallel to the plane of the membrane (``cholesteric" phase with $\Psi = 0$). Though the introduction of $\Psi$ provides a useful and predictive theory, it is not clear how it could be measured.  The KM theory also introduces terms in the free energy that are not directly present in our theory: one measuring the energy cost of surface curvature and two providing a favored relative orientation of the surface normal $\veh N$ and director $\ve n$ at the top and bottom membrane surfaces. However, the term proportional to $-h/(t+b_0)$ in the Eq.~\ref{eqn:fcartesian} version of our theory provides a preference of $\theta=0$, i.e., the director prefers to be parallel to the layer normal. More generally, the Lagrange multiplier term in Eq.~\ref{eqn:f} provides this preference. Naturally, the KM theory has more free parameters than the five of our theory: depletant size, Frank elastic constant, twist wavenumber amplitude and transition temperature, and virus birefringence (Table~\ref{tab:1}). In spite of these differences between the two theories, they share some common features: They both employ the Frank free energy with a term favoring twist to describe the energetics of director deformations, and they both introduce a term favoring $h = t\cos\theta$ (when $b_0$ can be ignored in our theory) with a coefficient ($\cos \theta$ in our case and $|\Psi|^2$ in the KM case) that vanishes at a $\pi$-wall when $\theta = \pi/2$, importantly allowing $h$ to differ from $t\cos\theta$ with no direct energy cost at that point.

KM pursues a different approach to boundary conditions than we do.  They impose the condition $\theta(R)=\pi/2$ at the free edge of a circular membrane, whereas we allow the Euler-Lagrange equations of our theory to set the conditions on $\theta$ and $h$ at the edge.  As a result, we are able to capture the edge profiles of small membranes whose rods are clearly not parallel to the membrane. Presumably, KM theory is amenable to the same approach and could thus calculate edge profiles of small membranes. KM also view the membrane thickness at the $\pi$-wall as a boundary condition determined by experiment, whereas it is a prediction of our model once physical parameters have been set.

The KM fits to edge and $\pi$-wall retardance data reported in Refs.~\cite{Zakhary:2014de,Kaplan:2014cz} (e.g., Fig.~6 of Ref.~\cite{Kaplan:2014cz}) are seemingly better than the fits in Figs.~\ref{fig:edgeret}(b) and \ref{fig:wallret}(b) from our theory.  It should be noted, however, that we use one set of parameters to fit data from all membrane radii, whereas the KM fits only consider data from a single radius. Our fit to individual profiles are as good as those of the KM theory.

\subsection{\label{ssec:ripple}Edge ripple fluctuations}

Our free energy Eq.~\ref{eqn:fcartesian} can also be used to investigate edge ripple fluctuations of large membranes. First, we minimize the free energy over $h(x)$ and $\theta(x)$ to obtain the profile for the unperturbed membrane edge. We then introduce a small edge ripple $u(y)$ with corresponding tangent angle $\alpha(y) \equiv \partial_y u(y)$. We assume that the edge profile completely propagates into the membrane interior, so $h(x,y)=h(x-u(y))$, and that the rod tilt follows the tangent of $u(y)$, so the nematic director changes from $\ve n(x)=\sin\theta(x)\veh y + \cos\theta(x)\veh z$ to $\ve n(x,y)=\sin\alpha(y)\sin\theta(x-u(y))\veh x + \cos\alpha(y)\sin\theta(x-u(y))\veh y + \cos\theta(x-u(y))\veh z$ [for a schematic of the ansatz, see Fig.~\ref{fig:tension}(a)]. We have to rederive the depletion and Frank terms in Eq.~$\ref{eqn:fcartesian}$ to allow for gradient terms in the $y$-direction (expression not shown here). We expand the ripple tangent angle in Fourier components $\alpha_p$:
\begin{equation}
	\alpha(y) = \sum_p \sqrt{\frac{2}{L_y}} \alpha_p \cos py.
	\label{eqn:fourier}
\end{equation}
$p$ is the ripple wavenumber~\cite{Gibaud:2012cf}. With the help of $\alpha_p = p u_p$, where $u_p$'s are Fourier components for $u(y)$, we can write the free energy in terms of the small $\alpha_p$'s. The free energy relative to the state without ripples becomes
\begin{equation}
	\frac{\Delta F}{L_y} = \frac{1}{2} \sum_p \left(\gamma[h,\theta] + \kappa[h,\theta] p^2\right) \alpha_p^2 + \order(\{\alpha_p\}^4),
	\label{eqn:deltaf}
\end{equation}
which describes a 1D interface with effective line tension $\gamma$ and line bending modulus $\kappa$. They are given by
\begin{eqnarray}
	\gamma[h,\theta] &=& 2na\kT\left[\int_0^{L_x}\dd x \frac{(\partial_x h)^2}{\sqrt{1+(\partial_x h)^2}} + h(0)\right] \nonumber\\
	&& {}+ 2K \int_0^{L_x} \dd x\, h\left[(\partial_x\theta)^2 + q\partial_x\theta\right],
	\label{eqn:gamma}\\
	\kappa[h,\theta] &=& 2K \int_0^{L_x}\dd x\, h\sin^2\theta.
	\label{eqn:kappa}
\end{eqnarray}
At thermal equilibrium, the ripple tangent angle components take the equipartition values
\begin{equation}
	\avg{\alpha_p^2} = \frac{\kT}{\gamma + \kappa p^2}.
	\label{eqn:avgalpha}
\end{equation}

Note that the term proportional to the chiral twist wavenumber $q$ in Eq.~\ref{eqn:gamma} is negative for $\partial_x\theta < 0$. The variation of its magnitude with temperature [$q(T)$ is temperature-dependent] is the theoretical basis for the chiral control of line tension presented in Fig.~\ref{fig:tension} and Ref.~\cite{Gibaud:2012cf}. All the other terms are positive-definite, so this term must be responsible for the line tension becoming negative at low temperatures, leading to the starfish instability. It is analogous to the chiral line tension term in the theory of Langmuir-Blodgett films, which if sufficiently negative, can drive an instability transition from a circular film to one with similarly extended arms~\cite{Pettey:1999jx}.

Next we investigate the dynamics of ripple fluctuations. We view the membrane edge as an effective 1D viscous fluid described by the ripple profile $u(y,t)$, which can vary with time. We estimate the Reynolds number of this motion to be very small ${\sim}10^{-6}$--$10^{-4}$, so the ripple velocity $v = \partial_t u$ obeys overdamped 1D hydrodynamics:
\begin{equation}
	{-}\eta_\textrm{1D}\partial_y^2 v = f_\textrm{ext} = f_\textrm{drag}[v] - \deldel{\mathcal{H}_T}{u}.
	\label{eqn:navier}
\end{equation}
$\eta_\textrm{1D}$ is the 1D edge viscosity and $f_\textrm{drag}[v]$ is the viscous drag force per unit length arising from membrane edge motion relative to the bulk solvent~\cite{chaikinlubensky}. Different models of membrane-fluid interactions lead to different expressions for $f_\textrm{drag}$; we see in Appendix~\ref{sec:drag} that it can be largely ignored for ripple wavenumbers probed by our experiments. In other words, dissipation of ripple excitations occurs mainly through the membrane rather than surrounding solvent since the membrane has much higher viscosity. Using $\mathcal{H}_T = \Delta F/L_y - \sum_p f_p u_p$ for the total Hamiltonian density, where $\Delta F/L_y$ is given by Eq. \ref{eqn:deltaf} and the $f_p$'s are an external field formally included to calculate the response function, we obtain:
\begin{equation}
	\eta_\textrm{1D}p^2\partial_t u_p = -\left(\gamma p^2 + \kappa p^4\right)u_p + f_p.\nonumber
	\label{eqn:navier2}
\end{equation}
This leads to the response function
\begin{equation}
	\chi_{u_pu_p}^{-1}(\omega) = \parpar{f_p(\omega)}{u_p(\omega)} = -\ii \omega \eta_\textrm{1D} p^2 + \gamma p^2 + \kappa p^4.\nonumber
	\label{eqn:chi}
\end{equation}
The fluctuation-dissipation theorem gives the autocorrelation function:
\begin{equation}
	S_{u_pu_p}(\omega) = \frac{2\kT}{\omega} \mathop\mathrm{Im} \chi_{u_pu_p} = \frac{2 \kT}{\eta_\textrm{1D} p^2}\frac{1}{\omega^2 + {\omega_p}^2},\nonumber
	\label{eqn:s}
\end{equation}
where 
\begin{equation}
	\omega_p \equiv \frac{\gamma + \kappa p^2}{\eta_\textrm{1D}}
	\label{eqn:omega}
\end{equation}
is the autocorrelation decay rate. Indeed, temporal ripple angle autocorrelations are given by
\begin{equation}
	\avg{\alpha_p(t)\alpha_p(0)} = \int \frac{\dd\omega}{2\pi} \ee^{\ii\omega t} p^2 S_{u_pu_p}(\omega) = \avg{\alpha_p^2} \ee^{-\omega_p t},
	\label{eqn:alphat}
\end{equation}
with $\langle \alpha_p^2 \rangle$ in Eq.~\ref{eqn:avgalpha}.

\section{\label{sec:exp}Experimental methods}

As model rod-like colloids, we use two strains of the filamentous \emph{fd} bacteriophage: wildtype (wt) and the Y21M mutant~\cite{Barry:2009uv}. As compared to \emph{fd}-wt, \emph{fd}-Y21M has a single point mutation in which the 21st amino acid of the major coat protein is changed from tyrosine (Y) to methionine (M). Both viruses have the same contour length, \SI{880}{\nm}, and diameter, \SI{6.6}{\nm}; their persistence lengths are \SI{2.8}{\um} for \emph{fd}-wt and \SI{9.9}{\um} for \emph{fd}-Y21M. They form cholesteric phases with opposite handedness: \emph{fd}-wt forms left-handed cholesterics whereas \emph{fd}-Y21M forms right-handed cholesterics. Finally, the chirality of \emph{fd}-wt is temperature-sensitive whereas the chirality of \emph{fd}-Y21M is temperature-independent~\cite{Gibaud:2012cf}.
 
Both viruses are synthesized using standard biological protocols~\cite{maniatis}. After synthesis, we observe a small portion of viruses that are very long---two and three times the nominal length of the virus. We fractionated the viruses through the isotropic-nematic phase transition; only the isotropic fraction, enriched in nominal-length viruses, is kept for this work~\cite{Gibaud:2012cf}. These monodisperse viruses are then dispersed with concentration $c_\textrm{virus} = \SI{1}{\mg\per\mL}$ in \SI{20}{\milli\molar} Tris buffer at pH 8.0 and \SI{100}{\milli\molar} NaCl. Dextran (\SI{500}{\kilo\dalton}, Sigma-Aldrich) is used as a depletant agent.
 
Samples are prepared between glass cover slides and coverslips in homemade chambers. A layer of unstretched Parafilm is used as a spacer. Slides are coated with polyacrylamide brushes to prevent nonspecific binding of the viruses with the glass slides and to suppress the depletion interaction between viruses and the glass walls~\cite{Lau:2009gw}. Samples are made airtight using UV-treated glue (Norland Optical). Microscopy observations were performed with the inverted microscope Nikon Eclipse Ti equipped with an oil immersion objective (1.3 NA, 100x Plan-Fluor). Data is acquired using a cooled CCD camera (Andor Clara) for low acquisition rates (below \SI{50}{\Hz}) and Phantom v9.1 (Vision Research) for fast acquisition rates (above \SI{1000}{\Hz}).
 
Sample temperature is tuned between 4 and \SI{60}{\celsius} with a homemade Peltier module equipped with a proportional-integral-derivative temperature controller (ILX Lightwave LPT 5910). The temperature-controlling side of the Peltier device is attached to a copper ring fitted around the microscope objective, which heats or cools the sample through the immersion oil. A thermistor, placed in the copper ring adjacent to the sample, enabled the proportional-integral-derivative feedback necessary to adjust the temperature. Excess heat is removed using a constant flow of room-temperature water. Such a device allows us to trigger the starfish instability as shown in Fig.~\ref{fig:images}(e).
 
The local tilt of the rods with respect to the optical axis of the microscope is determined using an LC-Polscope (Cambridge Research and Instrumentation)~\cite{Oldenbourg:2011kd}. LC-PolScope produces images in which the intensity of each pixel is the local retardance $D$ of the membrane. Such images can be quantitatively related to the tilting of the rods away from the membrane normal (the $z$-axis in Fig.~\ref{fig:images}). Rods in the bulk of a membrane are aligned along the $z$-axis, and LC-PolScope images appear black in that region. By contrast, for sufficiently large membranes, the bright birefringent ring along the membrane’s periphery indicates local rod tilting as shown in Fig.~\ref{fig:edgeret}(a). In Fig.~\ref{fig:wallret}(a), the LC-PolScope image of a $\pi$-wall indicates that the structure contains twist.

The time-independent analysis of thermal edge ripple fluctuations with DIC optical microscopy yields the line tension and the bending rigidity of the edge~\cite{safran,Gibaud:2012cf}. The acquisition is performed at \SI{1}{\Hz} so that the edge fluctuations are decorrelated. Intensity profile cuts along the perpendicular to the edge are fitted by a Gaussian and yield the conformation of the edge with subpixel accuracy. Each conformation is described in terms of the Fourier amplitudes $\alpha_p$ (Eq.~\ref{eqn:fourier}). Averaging over a sufficient number of uncorrelated images gives a fluctuation spectrum as shown in Fig.~\ref{fig:tension}(b), where the mean-square amplitude $\avg{\alpha_p^2}$ is plotted as a function of the wavenumber $p$. The dynamical analysis of thermal edge ripple fluctuations with DIC optical microscopy yields the autocorrelation decay timescale. The acquisition is performed at \SI{3000}{\Hz}. The autocorrelation decay timescale $1/\omega_p$ at a given wavenumber $p$ is obtained by fitting the temporal autocorrelation function of the Fourier amplitudes by a simple exponential (Eq.~\ref{eqn:alphat}). Measurements over a sufficiently long time give $1/\omega_p$ as a function of $p$ as shown in Fig.~\ref{fig:tension}(c).
 
Colloidal membranes can be manipulated using optical tweezers. The laser tweezers setup is built around an inverted Nikon TE-2000 microscope. A \SI{1064}{\nm} laser beam (Compass 1064, Coherent) is projected onto the back focal plane of an oil-immersion objective (Plan Fluor 100x, NA = 1.3) and subsequently focused onto the imaging plane. Using custom LabVIEW software, multiple trap locations were specified and used to stretch and manipulate membranes. Above \SI{2}{\watt} of laser power, one can rip off smaller membranes from a larger membrane to produce membranes between 0.5 to \SI{5}{\um} in diameter. This technique is used to study small membranes as shown in Fig.~\ref{fig:images}(c).

\section{\label{sec:dis}Discussion}

The microscopic building components required for assembly of colloidal membranes are monodisperse rod-like viruses, non-adsorbing dextran polymer, and polyelectrolytes to screen electrostatic repulsion. Despite their relative simplicity, these building blocks can assemble into a myriad of complex structures. Our theory demonstrates how their rich properties can emerge from hard rods and depletants through three simple entropic considerations: depletant excluded volume, rod fluctuations perpendicular to the membrane, and rod twisting as described by the Frank free energy. For example, the curved membrane edge with chiral rods arises from the competition between depletion, which prefers a circular vertical edge profile, and Frank elasticity, which prefers an exponential edge twisting profile (Fig.~\ref{fig:edges} and Appendix~\ref{sec:breaking}). If depletion is strong enough compared to the Frank contribution, achiral virus mixtures will also form twisted membranes through spontaneous symmetry breaking. Furthermore, our theory predicts that smaller membranes, with less distance over which rods can twist, have more squared-off edge profiles; this prediction was verified by additional experimental data (Fig.~\ref{fig:edgeret}). Decreasing the temperature increases the preferred twist wavenumber and consequently decreases the energy of the membrane edge, where the twist is greatest. Thus, ripple fluctuations, which lengthen the membrane edge, increase in amplitude (Fig.~\ref{fig:tension}). Eventually, at low enough temperatures, edges are energetically preferred and ripples are stabilized in a twisted starfish configuration (Fig.~\ref{fig:starfish}). Besides explaining the properties of the membrane edge, our theoretical model can also explain the structure of $\pi$-walls. Along membrane edges, a high depletion concentration strongly suppresses rod fluctuations perpendicular to the membrane (Fig.~\ref{fig:edges}). Along the middle of $\pi$-walls, however, large rod fluctuations, which can be interpreted experimentally as rod stacking, decreases the depletants' excluded volume and are thus favored (Fig.~\ref{fig:walls}). This stack of rods with finite thickness physically connects the two partially coalesced membranes and, through depletion, keeps them together.

All variables in our theory have direct physical meaning. We directly manipulate two of these parameters---temperature and depletant concentration---and measure several independent physical properties---membrane retardance (Figs.~\ref{fig:edgeret} and \ref{fig:wallret}) and edge fluctuation spectra (Fig.~\ref{fig:tension}). Theoretical calculations of these properties demonstrate respectable agreement with experimental measurements while using physically reasonable parameter values (Table~\ref{tab:1}). We use values for the hard-sphere depletant size $a$ and \emph{fd} virus birefringence $\Delta n$ that are within ${\sim}25\%$ of the reported values. We require $q(T)$ to have its measured square-root behavior. The Frank elastic constant $K$ and preferred twist wavenumber $q(T)$ are ${\sim}5$ times larger than the values measured from viruses dispersed in a bulk cholesteric phase without any depletant. However, $K$ and $q$ depend strongly upon the virus concentration~\cite{Dogic:2000tp}; membranes condensed by depletants have a higher virus concentration than cholesteric suspensions do and thus should have higher $K$ and $q$.

Our theory uses a number of assumptions and simplifications. We study the membrane in the continuum limit with only two coarse-grained degrees of freedom. We ignore rod-rod interactions other than those implicit in the phenomenological Frank free energy, whose moduli are assumed to be equal and temperature-independent. Rod fluctuations perpendicular to the membrane do not directly increase the membrane's volume in the simple manner assumed, and while these fluctuations are most important at large rod angles $\theta\approx\pi/2$, their energetic cost (Eq.~\ref{eqn:frod}) was calculated in the small rod angle, small fluctuation amplitude limit. In addition, the retardance formula was derived for a material of constant thickness and optical axis, which does not apply to our membranes. We assume a simple ripple ansatz to calculate edge fluctuation spectra, but the actual ripples may have a different configuration with lower energy. Yet, despite all of these approximations, our model can match experimental results with quantitative accuracy, indicating that it still has value in describing and elucidating properties of colloidal membranes. 

The role of depletion and other hard-core interactions in colloidal systems has been vigorously investigated from many perspectives. Direct excluded volume minimization was used to study depletion-driven helix formation in elastic tubes~\cite{Snir:2005fc}. Effective entropic potentials between two anisotropic colloidal particles have been calculated in depth~\cite{vanAnders:2014cj} and have been used to explain various self-assembly processes~\cite{Sacanna:2011dd,Sacanna:2013hw,Kraft:2012cc,Ashton:2013cr}. Free-volume theory and theories based on pair distribution functions have probed the depletion-induced phase separation of colloidal species and have provided relatively sophisticated expressions for effective interfacial tensions~\cite{Lekkerkerker:2011ig,Oversteegen:2005kl,Aarts:2004go,Vrij:1997je}. However, to our knowledge, the depletion interaction has never appeared before as an effective surface tension of magnitude $na\kT$ explicity. Our system admits this expression because there is near-complete phase separation between the colloids and the depletants and because depletion is strong enough to fix the membrane volume in the continuum limit. Our $na\kT$ surface tension can be related to scaling arguments near the coexistence line in Flory-Huggins-de Gennes theory, which proposes an interfacial tension proportional to $\kT/l_\textrm{inter}^2$, where $l_\textrm{inter}$ is the thickness of the interface between colloid-rich and colloid-poor phases~\cite{degennes,deHoog:1999io,Lekkerkerker:2011ig}. Taking this thickness approximately to be the equilibrium rod height fluctuation amplitude $b_0$ calculated in our theory, our surface tension expression agrees with that obtained by scaling: $\kT/l_\textrm{inter}^2 \sim \kT/b_0^2 \sim na\kT$. Moreover, the ability of our model to quantitatively match and predict experimental results supports the validity of our expression, which may guide the design of other colloidal systems whose surface tension can be easily tuned by changing depletant concentration, depletant size, or temperature.

\begin{acknowledgments}
We are grateful for helpful discussions with Robert A. Pelcovits. We also acknowledge financial support from the National Science Foundation through grants DMR-1104707 (to L. K. and T. C. L.) and MRSEC-1206146 and DMR-0955776 (to Z. D.), and from the Agence Nationale de la Recherche through grant ANR-11-PDOC-027 (to T. G.).
\end{acknowledgments}

\appendix

\section{\label{sec:breaking}Spontaneous chiral symmetry breaking at membrane edges}

As discussed in Section~\ref{ssec:memb} and demonstrated in Fig.~\ref{fig:edges}(c), rod height fluctuations are strongly suppressed in membrane edge configurations. We can simplify the free energy (Eq.~\ref{eqn:fcartesian}) by enforcing $h = t\cos\theta + b_0$ and approximating $b_0 = 0$. The free energy can then be expressed in terms of $\theta$ only. In a dimensionless form with $\tilde x \equiv x/t$, $\tilde L_x \equiv L_x/t$, $\tilde q \equiv qt$, and $\tilde F \equiv F/2nat\kT L_y$,
\begin{eqnarray}
	\tilde F &=& \int_0^{\tilde L_x} \dd \tilde x \left(\sqrt{1+\sin^2\theta\,(\partial_{\tilde x}\theta)^2} - \cos\theta\right) + \cos\theta(0) \nonumber\\
	&& {}+ \frac{k}{2} \int_0^{\tilde L_x} \dd \tilde x\,\cos\theta\,(\partial_{\tilde x}\theta)^2 - k\tilde q \sin\theta(0).
	\label{eqn:ftheta}
\end{eqnarray}
To investigate the onset of twist, we expand this free energy for small $\theta$. To third order, the first integral of the Euler-Lagrange equation gives
\begin{equation}
	\sqrt{k}\partial_{\tilde x}\theta = -\theta + \frac{12-5k}{24k}\theta^3. \nonumber
\end{equation}
This equation at $\tilde x = 0$ can be combined with the variational boundary condition
\begin{equation}
	k\partial_{\tilde x}\theta(0) = -k\tilde q - \theta(0) + \tilde q \theta^2(0) + \frac{3-k}{3k}\theta^3(0) \nonumber
\end{equation}
to obtain $\theta(0)$. We first consider $\tilde q = 0$, so $\tilde F$ has chiral symmetry. We find a twist solution when $k < k_c = 1$, where
\begin{equation}
	\theta(0) \approx \pm \sqrt{\frac{4}{3}(1-k)}
	\label{eqn:theta01}
\end{equation}
close to the critical point. When $k > 1$, only the trivial $\theta = 0$ solution exists. If we allow a small nonzero $\tilde q$ to break the chiral symmetry, a twist solution appears above $k_c$:
\begin{equation}
	\theta(0) \approx \frac{k \tilde q}{\sqrt{k} - 1}.
	\label{eqn:theta02}
\end{equation}
We can integrate the Euler-Lagrange equation to leading order and obtain
\begin{equation}
	\theta(\tilde x) \approx \theta(0)\exp(-\tilde x/\sqrt{k}).
	\label{eqn:theta}
\end{equation}
$\sqrt{k}t$ acts as a twist penetration depth $l_\textrm{twist}$ in analogy to smectic phases. Free energy calculations confirm that the twist solutions are favored whenever they exist. Thus, when $q=0$, the phase transition at the $k_c = 1$ critical point is second-order and spontaneously breaks chiral symmetry. Above $k_c$, there is a critical second-order line at $q=0$.

We also investigate the edge profile when $k \ll 1$. It is more convenient to write the free energy (Eq.~\ref{eqn:ftheta}) in terms of $\tilde h \equiv h/t = \cos\theta$:
\begin{eqnarray}
	\tilde F &=& \int_0^{\tilde L_x} \dd \tilde x \left(\sqrt{1+(\partial_{\tilde x}\tilde h)^2} - \tilde h\right) + \tilde h(0) \nonumber\\
	&& {}+ \frac{k}{2} \int_0^{\tilde L_x} \dd \tilde x \frac{\tilde h(\partial_{\tilde x}\tilde h)^2}{1-\tilde h^2} - k\tilde q \sqrt{1-\tilde h^2(0)}.
	\label{eqn:ftildeh}
\end{eqnarray}
We chose the sign of the square-root in the last term assuming $\theta > 0$, so this expression applies for $q > 0$. If $q < 0$, then $\theta < 0$ configurations have lower energy and we should choose the opposite sign. The first integral of the Euler-Lagrange equation gives
\begin{equation}
	0 = \frac{1}{\sqrt{1+(\partial_{\tilde x}\tilde h)^2}} - \tilde h - \frac{k}{2} \frac{\tilde h (\partial_{\tilde x}\tilde h)^2}{1-\tilde h^2}. \nonumber
\end{equation}
This equation at $\tilde x = 0$ can be combined the variational boundary condition
\begin{eqnarray}
	0 &=&  \frac{\partial_{\tilde x}\tilde h(0)}{\sqrt{1+(\partial_{\tilde x}\tilde h(0))^2}} - 1 \nonumber\\
	&& {}+ k \frac{\tilde h(0) \partial_{\tilde x}\tilde h(0)}{1-\tilde h^2(0)} - k\tilde q \frac{\tilde h(0)}{\sqrt{1-\tilde h^2(0)}} \nonumber
\end{eqnarray}
to obtain a twist solution as a power series in $k$:
\begin{equation}
	\tilde h(0) \approx \sqrt{\frac{27}{32}k} - \frac{9}{8} k\tilde q.
	\label{eqn:tildeh0}
\end{equation}
Solving the Euler-Lagrange equation with $k = 0$ yields a circular profile
\begin{equation}
	\tilde h(\tilde x) \approx \begin{cases} \sqrt{2\left(\tilde x + \frac{\tilde h^2(0)}{2}\right) - \left(\tilde x + \frac{\tilde h^2(0)}{2}\right)^2} & \tilde x \leq 1 - \frac{\tilde h^2(0)}{2} \\ 1 & \tilde x > 1 - \frac{\tilde h^2(0)}{2}. \end{cases}
	\label{eqn:tildeh}
\end{equation}
However, since $\cos\theta(0) = \tilde h(0) \ll 1$, the coupling in $F_\textrm{rod}$ may be weak. Calculations using the full free energy should be performed to check if $h = t\cos\theta$ is a valid assumption. 

\section{\label{sec:drag}Estimation of solvent drag during ripple fluctuations}

Here we estimate the dissipative forces exerted by the solvent on the membrane as it undergoes ripple fluctuations. We approximate the membrane as an infinite 2D fluid and apply the analysis of \cite{Lubensky:1996cg}, who consider the drag force exerted by a subfluid of depth $d$ below the fluid plane. For a velocity field $\ve v = v\veh x$ with wavevector $\ve p = p\veh y$, the drag per unit area is $g_\textrm{drag} = -\eta_\textrm{s} p \coth{(pd)}\,v$, where $\eta_\textrm{s}$ is the subfluid viscosity. We estimate a solvent depth $d \sim \SI{0.3}{\um}$ under the membrane where the polymer brush lies. The fluid above the membrane plane exerts much less drag because $\coth{pd}$ is a monotonically decreasing function of $d$, so we ignore it. Assuming that an effective width $l_x \sim l_\textrm{twist} \sim t$ of the membrane edge moves during the ripple fluctuations, the drag force per unit length is approximately $f_\textrm{drag} = \int \dd x\,g_\textrm{drag} \sim g_\textrm{drag}l_x$. This force modifies the fluctuation autocorrelation decay constant (Eq.~\ref{eqn:omega}) to
\begin{equation}
	\omega_p = \frac{\gamma + \kappa p^2}{\eta_\textrm{s} l_x \coth (pd)/p + \eta_\textrm{1D}}.\nonumber
\end{equation}
With $\eta_\textrm{s} \approx \SI{3}{\milli\pascal\s}$ from \cite{Cush:1997ky} and $\eta_\textrm{1D} \approx \SI{300}{\milli\pascal\s\square\um}$ from this work, this change would increase the calculated values of $1/\omega_p$ in Fig.~\ref{fig:tension}(c) at low wavenumbers $p \lesssim \SI{0.3}{\per\um}$, but it would not significantly modify our fit value for $\eta_\textrm{1D}$. For example, at $p = \SI{0.3}{\per\um}$, $1/\omega_p$ would be increased ${\sim}$20\%. Moreover, since the measured values of $1/\omega_p$ do not show any increase at small $p$, this analysis may overestimate the drag force, a claim whose verification would require a much more sophisticated theory that better captures the ripple geometry and motion.

\bibliography{refs,refs-book}

\end{document}